\documentclass[journal,doublecolumn,singlespace,a4paper]{IEEEtran}

\ifCLASSINFOpdf
\else
\fi
\usepackage[dvips]{graphicx}
\usepackage{balance}
\usepackage{graphicx}
\usepackage{bm}
\usepackage{amsmath}
\usepackage{amsfonts}
\usepackage{amsthm}
\usepackage{amssymb}
\usepackage{MyMnSymbol}
\usepackage{color}
\usepackage{mathrsfs}
\usepackage[keeplastbox]{flushend}
\usepackage[utf8]{inputenc}
\usepackage{tikz}
\usetikzlibrary{arrows, calc, decorations.markings, positioning}
\usepackage{cite}
\usepackage{booktabs}
\usepackage{multirow}
\usepackage{amsmath}
\usepackage{amsthm}
\usepackage{algorithm}
\usepackage{algorithmic}

\usepackage{mathrsfs}
\usepackage{amsmath}
\usepackage{amssymb}
\usepackage{amsthm}
\usepackage{lineno}
\usepackage{color}
\usepackage[labelformat=empty,labelsep=none]{caption}
\usepackage{hyperref}
\usepackage{algorithm}
\usepackage{algorithmic}
\usepackage{indentfirst}
\usepackage{pifont}
\usepackage[normalem]{ulem}
\usepackage{epstopdf}
\usepackage{graphicx}
\usepackage{float}

\usepackage[flushleft]{threeparttable}
\ifCLASSOPTIONcompsoc
 \usepackage[caption=false,font=normalsize,labelfont=sf,textfont=sf]{subfig}
\else
 \usepackage[caption=false,font=footnotesize]{subfig}
\fi
\newtheorem{thm}{Theorem}

\usepackage{textcomp}
\usepackage{xcolor}
\def\BibTeX{{\rm B\kern-.05em{\sc i\kern-.025em b}\kern-.08em
    T\kern-.1667em\lower.7ex\hbox{E}\kern-.125emX}}

\usepackage{tikz}

\newdimen\satlevel
\newdimen\satdiameter
\newcommand{\satisfaction}[2][]{%
	\satdiameter=1.9ex\relax
	\ifcase#2\relax
	\satlevel=0pt\relax
	\or
	\satlevel=0.125\satdiameter
	\or
	\satlevel=0.25\satdiameter
	\or
	\satlevel=0.375\satdiameter
	\or
	\satlevel=0.5\satdiameter
	\fi
	\tikz[baseline=-0.3\satdiameter]{%
		\draw[#1] (0,0) circle (0.5\satdiameter);
		\fill[#1] (0,0) circle (\satlevel);
	}%
}


\begin{document}
%\linenumbers

\title{Design and Performance Evaluation of Joint Sensing and Communication Integrated System for 5G MmWave Enabled CAVs}

\author{Qixun~Zhang, ~\IEEEmembership{Member,~IEEE,}
        Xinna~Wang,
        Zhenhao~Li,
        Zhiqing~Wei, ~\IEEEmembership{Member,~IEEE}
        %Zhiyong~Feng, ~\IEEEmembership{Senior Member,~IEEE}
\thanks{Corresponding author is Q. Zhang. Q. Zhang, X. Wang, Z. Li and Z. Wei are with the Key Laboratory of Universal Wireless Communications Ministry of Education, Beijing University of Posts and Telecommunications, Beijing 100876, China (e-mail: zhangqixun@bupt.edu.cn, xinna\textunderscore wang@bupt.edu.cn, zhenhao918@bupt.edu.cn, weizhiqing@bupt.edu.cn).}
\thanks{This work was partly supported by National Natural Science Foundation of China (NSFC) (Grant No. 62022020) and National Key Research and Development Project (Grant No. 2020YFB1807600).}
\thanks{Manuscript submitted to Journal of Selected Topics in Signal Processing on Feb. 18, 2021; Major revision submitted on Jun. 18, 2021.}
}

\maketitle

\begin{abstract}
The safety of connected automated vehicles (CAVs) relies on the reliable and efficient raw data sharing from multiple types of sensors. The 5G millimeter wave (mmWave) communication technology can enhance the environment sensing ability of different isolated vehicles. In this paper, a joint sensing and communication integrated system (JSCIS) is designed to support the dynamic frame structure configuration for sensing and communication dual functions based on the 5G New Radio protocol in the mmWave frequency band, which can solve the low latency and high data rate problems of raw sensing data sharing among CAVs. To evaluate the timeliness of raw sensing data transmission, the best time duration allocation ratio of sensing and communication dual functions for one vehicle is achieved by modeling the $M/M/1$ queuing problem using the age of information (AoI) in this paper. Furthermore, the resource allocation optimization problem among multiple CAVs is formulated as a non-cooperative game using the radar mutual information as a key indicator. And the feasibility and existence of pure strategy Nash equilibrium (NE) are proved theoretically, and a centralized time resource allocation (CTRA) algorithm is proposed to achieve the best feasible pure strategy NE. Finally, both simulation and hardware testbed are designed, and the results show that the proposed CTRA algorithm can improve the radar total mutual information by 26$\%$, and the feasibility of the proposed JSCIS is achieved with an acceptable radar ranging accuracy within $\pm$ 0.25 m, as well as a stable data rate of 2.8 Gbps using the 28 GHz mmWave frequency band.
\end{abstract}


\IEEEpeerreviewmaketitle

\section{Introduction}

In recent years, the rapid development of artificial intelligence and mobile communication technology has brought revolutionary progresses for the autonomous driving vehicles (ADVs). Typical sensors in ADVs include the optical cameras, light detection and ranging (LiDAR), ultrasonic sensors, mmWave radars and so on. However, the detection performance of different sensors is vulnerable to the dynamic changing radio environment and the blockage of other vehicles. In order to ensure the safety of ADVs, the cooperation of multi-sensors in different vehicles via wideband communication technology is an effective way to extend the environment information sensing ability and facilitate vehicles' decision-making [1]. But, multiple sensors in ADVs can generate an excessive amount of raw data over several Gbps [2], leading to the efficient raw data sharing among vehicles a big problem.

It is envisioned that the Internet of Vehicles (IoV) and the 5G mmWave communications [3] would be a promising solution to tackle the aforementioned challenges, which may significantly improve the see-through ability of ADVs [4]. As a fast-developing technology, 5G cellular vehicle-to-everything (C-V2X) is able to support a higher data rate [5]. Besides, the information interaction between vehicles via 5G communication can effectively fill the blind area of sensing, achieving the high-accuracy information integration. Furthermore, vehicles mounted with in-vehicle sensors and V2X communication equipment to support autonomous driving applications are defined as the connected automated vehicles (CAVs) [6]. Considering the limited spectrum resources for 5G C-V2X, the idea of spectrum sharing technology [7] using joint sensing and communication design has attracted much attention recently, which may solve the problem of low latency raw sensing data sharing and improve the sensing ability of CAVs. A distributed networking protocol RadChat was proposed in [7] to mitigate the interference among FMCW based automotive radars by using radar and communication cooperation. And the results show that RadChat can significantly reduce radar mutual interference in single-hop vehicular networks in less than 80 ms.

From the perspective of simple dual-functional aggregation, the information from radar has been used to assist the beam alignment of 60 GHz mmWave communication through the Dedicated Short Range Communication (DSRC) [1]. Furthermore, the combination design of the automotive radar at 76.5 GHz and the mmWave communication at 65 GHz has been proposed for the radar-aided communication with separate antennas [8], which only considers the vehicle-to-infrastructure (V2I) scenario in different spectrum bands without considering the V2V scenario. Although these aforementioned schemes have the basic sensing and communication functionalities, they only focus on the simple coexistence issue of two systems, which lead to severe signaling overhead and hardware cost, and are consequently unable to realize the integrated system.

In terms of the sensing and communication integration perspective, the existing research works focus on the joint frame structure and waveform design. Pioneered by [9], a radar information measurement using mutual information (MI) has been proposed. Furthermore, the fusion of radar and communication can be realized by using the radio frequency convergence schemes [10]. Moreover, MIMO radar and the cellular base stations can coexist through spectrum sharing, where the radar waveform can be used as a pilot signal [11]. Second, for the joint frame structure design, the radar and communication signals can be transmitted in different time slots without causing the interference to each other based on the time division framework [12]. A novel time division duplex frame structure which is capable of unifying the radar and communication operations is proposed in [13], and the signal processing is divided into three stages to realize basic radar and communication operations. For the purpose of designing a joint waveform, one may embed communication data into the radar probing signal [14]. Alternatively, the communication preambles of IEEE 802.11ad standard can also be exploited to design radar waveforms [15], but there is no system-level design and simulation for the integrated waveform. By using spatial and time division methods, both sensing and communication stages can be switched periodically with a set of configured antennas in [16]. Above all, the radar and communication functions can be complementary to each other in the integrated system. However, the sharing of mmWave band between radar and communication dual functions has not been fully considered in the existing research works, and the performances of sensing and communication cannot be balanced effectively given the limited computation resources and hardware constraints. Therefore, the resource allocation problem needs to be taken into account for the design of radar and communication dual functions. In [17], the non-cooperative game based resource allocation method is proposed to optimize the sum rate of the self-backhaul dense mmWave cellular network, which can also be used in the vehicle network. But, the radar and communication integrated system faces the challenging problem of complex interactions of sensing and communication signal. As a proper metric to characterize the information freshness [18] among CAVs, the Age of Information (AoI) is defined as the freshness of the received information. The AoI performance of a $N$-hop network with time-invariant packet loss probabilities on each link is analyzed in [19]. However, the packet loss probability will change with the variation of the network state, which is particularly serious in the vehicle network. Besides, a method to minimize the expected AoI satisfying timely-throughput constraints is proposed for a single-hop wireless network with a number of nodes transmitting the time-sensitive information to a base station in [20]. The average AoI for stationary randomized and round-robin scheduling policies in a multi-source status update system is analyzed in [21], but these policies are not suitable for the timeliness data sharing in the vehicle network. The vehicular beacon broadcasting scheduling problem with respect to the AoI is investigated in [22]. However, these research works have not considered the problem of data freshness in the radar and communication integrated system.

To cope with these challenges, we propose a time-division based joint sensing and communication integrated system (JSCIS) for 5G mmWave enabled CAVs. To tackle the low latency and efficient raw sensing data delivery problem, we model the source-destination link for CAVs using $M/M/1$ queuing theory and analyze the packet loss probability and average AoI. And the best time duration allocation ratio of sensing and communication dual functions for one vehicle is achieved. For multiple CAVs, we propose the Centralized Time Resource Allocation (CTRA) algorithm based on the non-cooperative game to reduce the computational complexity of time resource allocation. To the best of our knowledge, we are the first to consider the time resource allocation problem for sensing and communication dual functions in the time-division based JSCIS. The main contributions of this paper are summarized as follows.

\begin{itemize}
\item A time-division based JSCIS for CAVs in the mmWave frequency band is proposed. And 5G New Radio (NR) based dynamic frame structure for JSCIS is designed, so that each vehicle can allocate the time duration ratio according to the service requirements. Based on the time-division frame structure and the interference analysis, the average AoI equation of single vehicle as a function of server load rate is derived using First-Come-First-Served (FCFS) $M/M/1$ queuing theory. We analyze the effects of different duration allocation ratios on the packet loss probability and the average AoI. The best time duration allocation ratio range for one vehicle is achieved through the analysis of the packet loss probability and the average AoI.
\item We establish the optimization problem with the radar total MI as the performance evaluation indicator, and further formulate the optimization problem as a non-cooperative game to tackle the resource allocation of multiple CAVs, and the feasibility and existence of the pure strategy Nash equilibrium (NE) is proved theoretically. To achieve the best NE, we design the CTRA algorithm with the low complexity and fast convergence speed, and derive a set of optimal allocation ratios for JSCIS.
\item To verify the feasibility of JSCIS and the performance of CTRA algorithm, both software simulation and hardware testbed are developed. The software simulation results denote that the best time duration allocation ratio of one vehicle can be achieved. For multiple CAVs, the software simulation results show that the proposed CTRA algorithm can improve the radar total mutual information by 26\% compared to conventional algorithms. And the proportional set of vehicle time allocation can be obtained. The hardware testbed results prove that the feasibility of the proposed JSCIS can be achieved with an acceptable radar ranging accuracy within $\pm$ 0.25 m, as well as a stable data rate of 2.8 Gbps using the 28 GHz mmWave frequency band.

\end{itemize}

The rest of this paper is organized as follows. In Section II, we present the system model and problem formulation. A novel time-division based resource allocation algorithm with non-uniform durations are proposed in Section III. In Section IV, both software simulation and hardware testbed are designed and the results are discussed in detail. Finally, Section V concludes this paper.

%第二部分：系统模型和问题建模
\section{System Model and Problem Formulation}

%第二章的总领性段落。
We consider a 5G mmWave enabled CAVs scenario with JSCIS design. In this section, we describe the scenario and the frame structure design for JSCIS. Based on the time-division JSCIS, both radar signal and communication signal interferences scenario are analyzed in detail, respectively. Furthermore, the source to destination link is modeled by $M/M/1$ queuing theory and the average Age of Information (AoI) for one vehicle is also defined, which is a function of server load rate under the packet loss probability. Finally, we establish a resource allocation optimization function to maximize the radar total MI considering multiple CAVs. The symbols and definitions are summarized in Table I.

\begin{table*}[t]
\centering
 \caption{ \label{parameter setting}Table I. Symbols and definitions.}
 \begin{center}
 \begin{tabular}{l l l}
 \hline
 \hline
  {Symbol} & {Definition}\\
  \hline
  {${N_s}$} & {Number of subframes in each scheduling cycle}\\
  {$N$} & {Number of vehicles}\\
  {${a_n}$} & {Normalized radar sensing duration series}\\
  {${I_i}$} & {Time allocation index of vehicle $i$}\\
  {${n^{r - com}}$} & {Communication interference from other vehicles to radar signal of vehicle $i$}\\
  {${n^{r - rad}}$} & {Radar interference from other vehicles to radar signal of vehicle $i$}\\
  {${n^{c - com}}$} & {Communication interference from other vehicles to communication signal of vehicle $i$}\\
  {${n^{c - rad}}$} & {Radar interference from other vehicles to communication signal of vehicle $i$}\\
  {$h_{i,i}^t$} & {Propagation gain for the radar $i$-target-radar $i$ path}\\
  {$h_{i,j}^t$} & {Propagation gain for the radar $j$-target-radar $i$ path}\\
  {${G_t}$} & {Antenna gain at transmitter}\\
  {${G_r}$} & {Antenna gain at receiver}\\
  {$\sigma _{i,i}^{RCS}$} & {Radar Cross Section from target to radar $i$}\\
  {$\sigma _{i,j}^{RCS}$} & {Radar Cross Section from radar $j$ to radar $i$}\\
  {${R_i}$} & {Distance from radar $i$ to target}\\
  {${R_j}$} & {Distance from radar $j$ to target}\\
  {${P_j}$} & {Transmit power of vehicle $j$}\\
  {$g_{i,i}^{ch - r}$} & {Fading gain of vehicle $j$ in communication duration to vehicle $i$ in radar duration}\\
  {$\gamma _i^{rad}$} & {SINR of vehicle $i$ in radar duration of ${\bar a_n} - {\bar a_{n - 1}}$}\\
  {$\gamma _i^{com}$} & {SINR of vehicle $i$ in communication duration of ${\bar a_n} - {\bar a_{n - 1}}$}\\

  \hline
  \hline
 \end{tabular}
 \end{center}
\end{table*}

%A.系统模型 fig1-eps-converted-to.pdf
\subsection{Scenario Description}
\begin{figure}[t]
\centering
\includegraphics[width=0.5\textwidth]{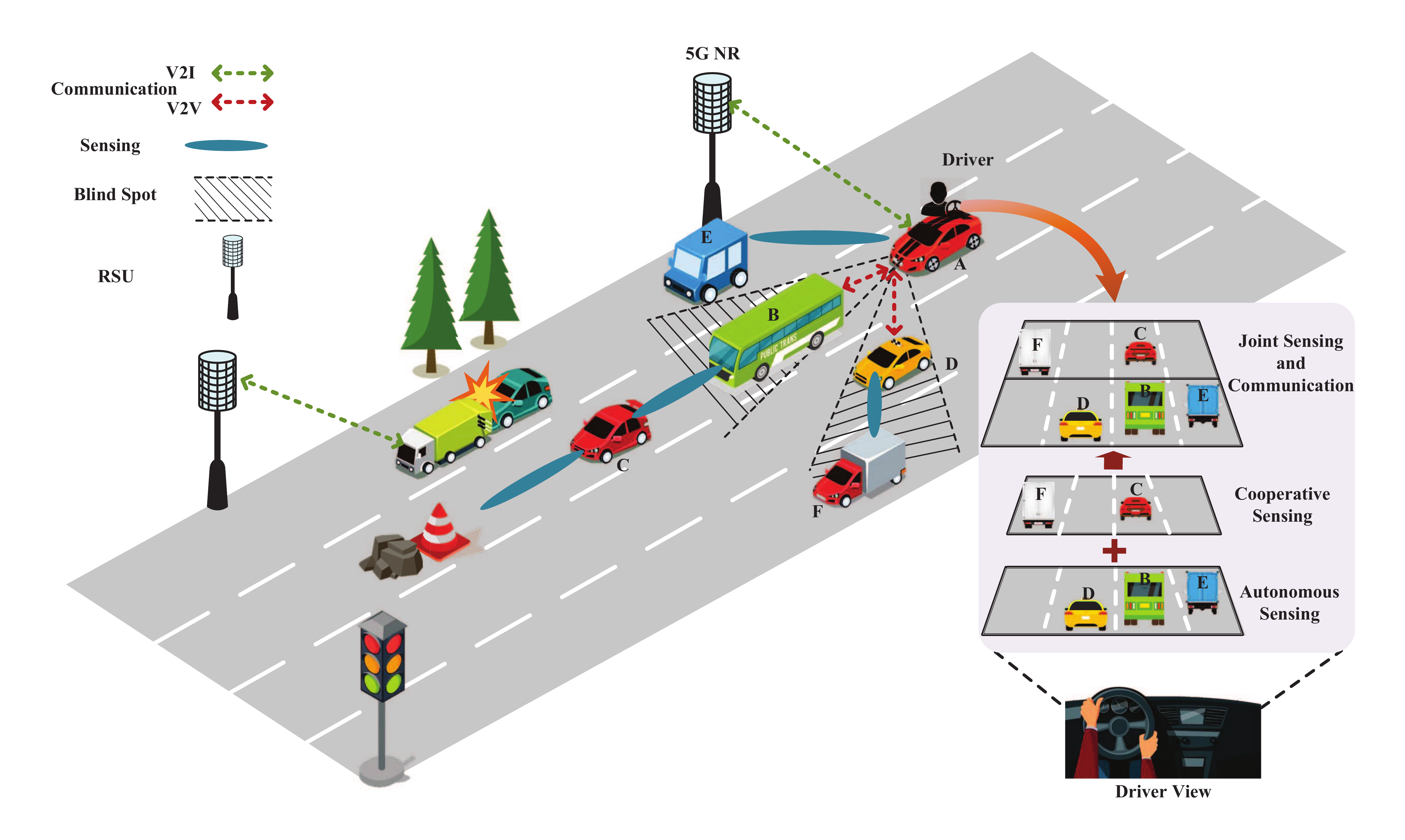}
\caption{Fig. 1. The sensing and communication integration scenario with multiple cooperative CAVs.}
\label{system scenario}
\end{figure}

The scenario of JSCIS for 5G mmWave enabled CAVs is shown in Fig. 1. Vehicle A can obtain the position information of vehicle B, D, and E through radar detection. However, vehicle A cannot detect vehicle C due to the blockage by vehicle B. By communicating with vehicle B, vehicle A can obtain the information about vehicle C. Each vehicle near vehicle A can use the mmWave link to directly communicate with vehicle A instantly, which can effectively reduce the transmission delay and assist vehicle A to be aware of the environment information nearby. In the same way, other vehicles outside the sensing distance of vehicle A or in the blind spot of the driver's view can also realize raw sensing information sharing using the mmWave communication links among vehicles. Vehicle A can also collect the information and communicate with the infrastructure on the roadside, while uploading the collected information and downloading other traffic or entertainment information. Therefore, vehicle A is considered as the central vehicle which is responsible for information processing in this paper based on ITU standard in [23], so as to effectively expand the sensing range. All time resource configuration sets, vehicle power configuration and channel state information (CSI) of all vehicles are known by vehicle A in the system. In addition, the resource allocation strategies adjusted by other vehicles can be sent back to the central vehicle A efficiently.

%B.帧结构设计
\subsection{Dynamic Frame Structure Design}
\begin{figure}[t]
\centerline{\includegraphics[width=0.5\textwidth]{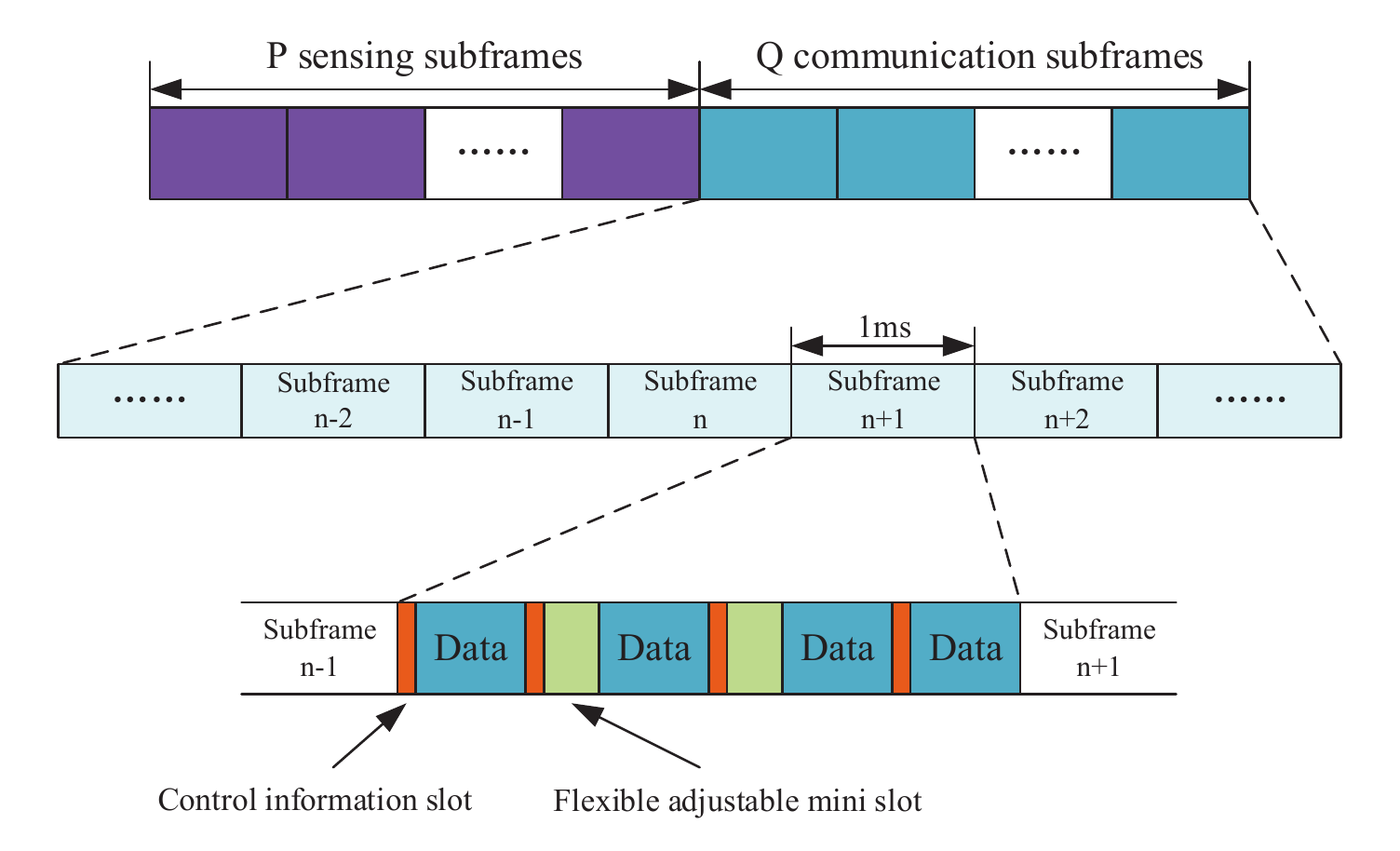}}
\caption{Fig. 2. The dynamic frame structure design of JSCIS.}
\label{frame structure}
\end{figure}

%引入5G帧结构的说明，为帧结构设计提供铺垫
Compared with the fixed 15kHz subcarrier interval in 4G LTE, 5G NR supports different subcarrier intervals [24]. For 5G NR V2X, the 15kHz, 30kHz and 60kHz subcarrier intervals are used to support the communication below 6GHz, while 60kHz and 120kHz can support the communication above 6GHz. Both transmitters and receivers can occupy more bandwidth resources in mmWave frequency bands. And the time delay will be considerably reduced by utilizing the longer subcarrier interval and shorter time slot. In the time domain, the length of a NR subframe is 1 ms, and the number of time slots in each subframe is integer.

%介绍本文中的帧结构设计
As shown in Fig. 2, the subframe is used as the minimum unit of resource allocation for JSCIS. We design a $P/Q$ adjustable frame format consisting of $P$ sensing subframes and $Q$ communication subframes. Considering the latency sensitive radar information requirements in the practical scenario, all the information detected by radar in different CAVs need to be transmitted within the next communication subframe.

%关于帧结构中通信子帧的灵活子帧结构设计及mini-slot
Taking into account different delay sensitivities of communication data, it is necessary to adopt a flexible time slots allocation algorithm in the communication subframe based on the 5G NR frame structure. Compared with 4G LTE system, the frame structure of 5G NR system is more flexible, where the mini-slot can meet the low-latency transmission requirements and support the small data packet transmission. The mini-slot can carry not only fewer OFDM symbols than ordinary time slots, but also the control information [24]. Therefore, we propose a flexible multi-slot combination strategy to realize the transmission of control information, sensing information, emergency information, and etc. When the emergency information needs to be transmitted, such as a car accident within a short distance ahead or pedestrians suddenly crossing the road, the adjustable mini-slots are used to achieve the low-latency transmission. For the ordinary data information transmission, the ordinary time slots are used. The low-latency and high-reliability information can be transmitted in the communication subframe through the proposed $P/Q$ adjustable frame format and mini-slot methods.

%C.系统性能评估指标
\subsection{System Performance Evaluation Indicators}

%关于雷达互信息的描述
To evaluate the performance of the proposed JSCIS, the performance metrics are proposed as follows. The radar system is designed to reduce the prior uncertainty of target by measuring and obtaining the target information. According to information theory, the larger the MI, the more information about the target can be obtained via measurements. Related researches have found that maximizing MI can effectively improve the target recognition ability of radar system in [25] and [26]. Therefore, to evaluate the performance of radar system, MI between the target response and the radar received signal is used as the metric for the radar estimation performance.

%雷达指标
\subsubsection{Radar Metric}

%关于雷达信号的表示
For a radar system, the baseband received signal model is established as
\begin{equation}\label{radar system}
y = s{h_{rad}} + n,
\end{equation}
where $y$ denotes the received signal, ${h_{rad}}$ is the propagation gain of radar response signal in the channel, and $n$ is the interference to the signal receiver (including internal thermal noise and other external signal interference). In order to compare the performance of radar and communication in the same dimension, we introduce the radar mutual information (MI) as the metric of radar signal design [27] to measure radar performance from the perspective of communication. In the communication system, the mutual information between X and Y refers to the amount of information in X containing Y or the amount of information in Y containing X. Therefore, the meaning of radar MI is the information amount of channel state information contained in the radar received signal. The goal in this paper is to maximize the utilization of MI. Therefore, we use MI to characterize the information amount of channel state information contained in the radar received signal in this paper, as $I\left( {y;h} \right)$ in
\begin{equation}\label{radarmutualinformation}
\begin{array}{l}
\mathop {I^{rad}}\left( {y;{h_{rad}}} \right) = H\left( y \right) - H\left( {y|{h_{rad}}} \right) = H\left( y \right) - H\left( n \right)\\
\;\;\;\;\;\;\;\;\;\;\;\;\;\;\;\;\;\;\;\;\; = \log \left( {1 + \frac{{s{h_{rad}}}}{n}} \right).
\end{array}
\end{equation}

%通信指标
\subsubsection{Communication Metric}
For a communication system, the channel capacity of an Additive White Gaussian Noise (AWGN) channel is defined as the maximum information transmission rate based on Shannon's theorem in [28] as
\begin{equation}\label{communication rate}
{C_{com}} = B{\log _2}\left( {1 + {\Gamma _{com}}} \right),
\end{equation}
where $B$ represents the system bandwidth, ${\Gamma _{com}}$ represents the signal to interference plus noise ratio (SINR). According to information theory, the channel capacity is the supremum of all achievable rates, which can also be defined as the maximum MI. Based on [29], the communication channel capacity during $T$ time can be expressed as
\begin{equation}\label{communication capacity}
{I^{com}} = T{\rm{\cdot}}B{\log _2}\left( {1 + {\Gamma _{com}}} \right).
\end{equation}

%D.问题建模
\subsection{Problem Formulation}

%简单介绍问题建模
The mmWave communication system uses the narrow beam to minimize the interference, making it possible for each vehicle to configure sensing and communication durations dynamically. In the proposed dynamic frame structure, each vehicle adopts a non-uniform frame which can adjust the time allocation for radar and communication according to different requirements. When the cross-interference between radar detection and communication transmission in CAVs is serious, this solution can also be degraded into a uniform time allocation strategy. Without loss of generality, we formulate and analyze the objective function of non-uniform time resource allocation scheme based on the dynamic frame structure.

%参数说明
The frame length of JSCIS is fixed, and the sensing duration of each vehicle can change dynamically according to the sensing and communication requirements. Assuming the number of vehicles in the multi-vehicle cooperative scenario is $N$, there are ${N_s}$ subframes in each scheduling cycle. We define ${a_n} \in A = [{a_1},{a_2}, \cdots ,{a_N}]$ as the radar sensing normalized duration series for all vehicles, where $n$ represents the time index. The set of values for ${a_n}$ is ${\Omega _t} = [\frac{1}{{{N_s}}},\frac{2}{{{N_s}}}, \cdots ,\frac{{{N_s} - 1}}{{{N_s}}}]$. The sensing duration of each vehicle is arranged in an ascending order and renumbered as ${\bar a_n} = \left[ {{{\bar a}_1},{{\bar a}_2}, \cdots ,{{\bar a}_N}} \right]$, where ${\bar a_1}$ and ${\bar a_N}$ are the smallest and the biggest normalized sensing durations, respectively. At the same time, two additional constants are introduced, ${\bar a_0} = 0$ and ${\bar a_{N{\rm{ + }}1}} = 1$. The index collection of these sensing duration series can be set as $I = [{I_1},{I_2}, \cdots ,{I_N}]$, where ${I_i}$ represents the time allocation index of vehicle $i$.

%干扰分析概要
Based on the time-division frame structure in Fig. 2, the interference analysis among vehicles is based on the synchronization among vehicles, which is feasible and supported by the Primary Synchronization Signal (PSS) and Secondary Synchronization Signal (SSS) within the 5G NR communication system in [38] and [42]. Therefore, the interference among vehicles in the radar and communication durations is analyzed as follows, based on the non-uniform time allocation scheme in JSCIS.

%1)雷达探测过程信号分析
\subsubsection{Signal Analysis During Radar Detection Process}

\begin{figure}[t]
\centerline{\includegraphics[width=0.5\textwidth]{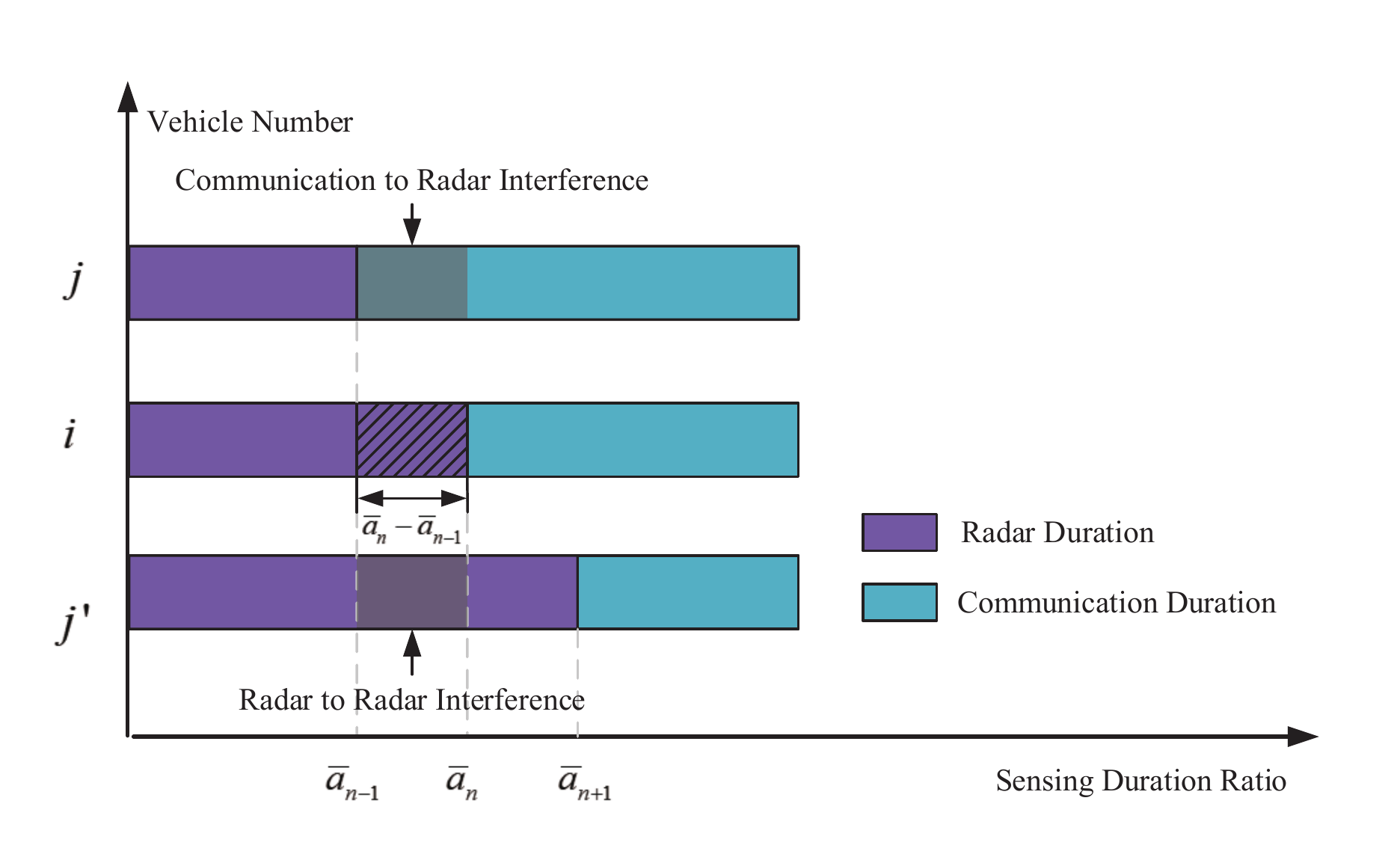}}
\caption{Fig. 3. Interference analysis in the radar duration of JSCIS.}
\label{Interference in radar duration}
\end{figure}

%图的解释
As shown in Fig. 3, in the radar duration ${\bar a_n} - {\bar a_{n - 1}}$ of vehicle $i$, and the time index $1 \le n \le {I_i}$, there are two types of interference, where the other vehicles are denoted by $j$ and $j'$. The radar signal of vehicle $i$ will be interfered by the communication signal of other vehicles when the index of other vehicles is smaller than $n$. And when the index of other vehicles is bigger than $n$, vehicle $i$ will experience the radar signal interference from other vehicles. Therefore, during the radar detection process, the radar received signal can be described as
\begin{equation}\label{radarsignal}
{y_{rad}} = {s_{rad}}{h_{rad}} + {n^{r - com}} + {n^{r - rad}} + n,
\end{equation}
where ${n^{r - com}}$ is the interference caused by communication signals from other vehicles to radar signals of vehicle $i$, and ${n^{r - rad}}$ is the interference caused by radar signals of other vehicles to radar signal of vehicle $i$.

%关于雷达信号传输增益的说明
There are two main paths for radar signal propagation, namely the reflection path and the refraction path. The corresponding path propagation gain is as follows
\begin{equation}\label{radar interference}
\left\{ \begin{array}{l}
h_{i,i}^t = \frac{{{G_t}{G_r}\sigma _{i,i}^{RCS}{\lambda ^2}}}{{{{(4\pi )}^3}R_i^4}},\\
h_{i,j}^t = \frac{{{G_t}{G_r}\sigma _{i,j}^{RCS}{\lambda ^2}}}{{{{(4\pi )}^3}R_i^2R_j^2}},
\end{array} \right.
\end{equation}
where $h_{i,i}^t$ represents the propagation gain for the radar $i$-target-radar $i$ path, and $h_{i,j}^t$ represents the propagation gain for the radar $j$-target-radar $i$ path. The detailed diagram is showed in Fig. 4.

\begin{figure}[t]
\centerline{\includegraphics[width=0.5\textwidth]{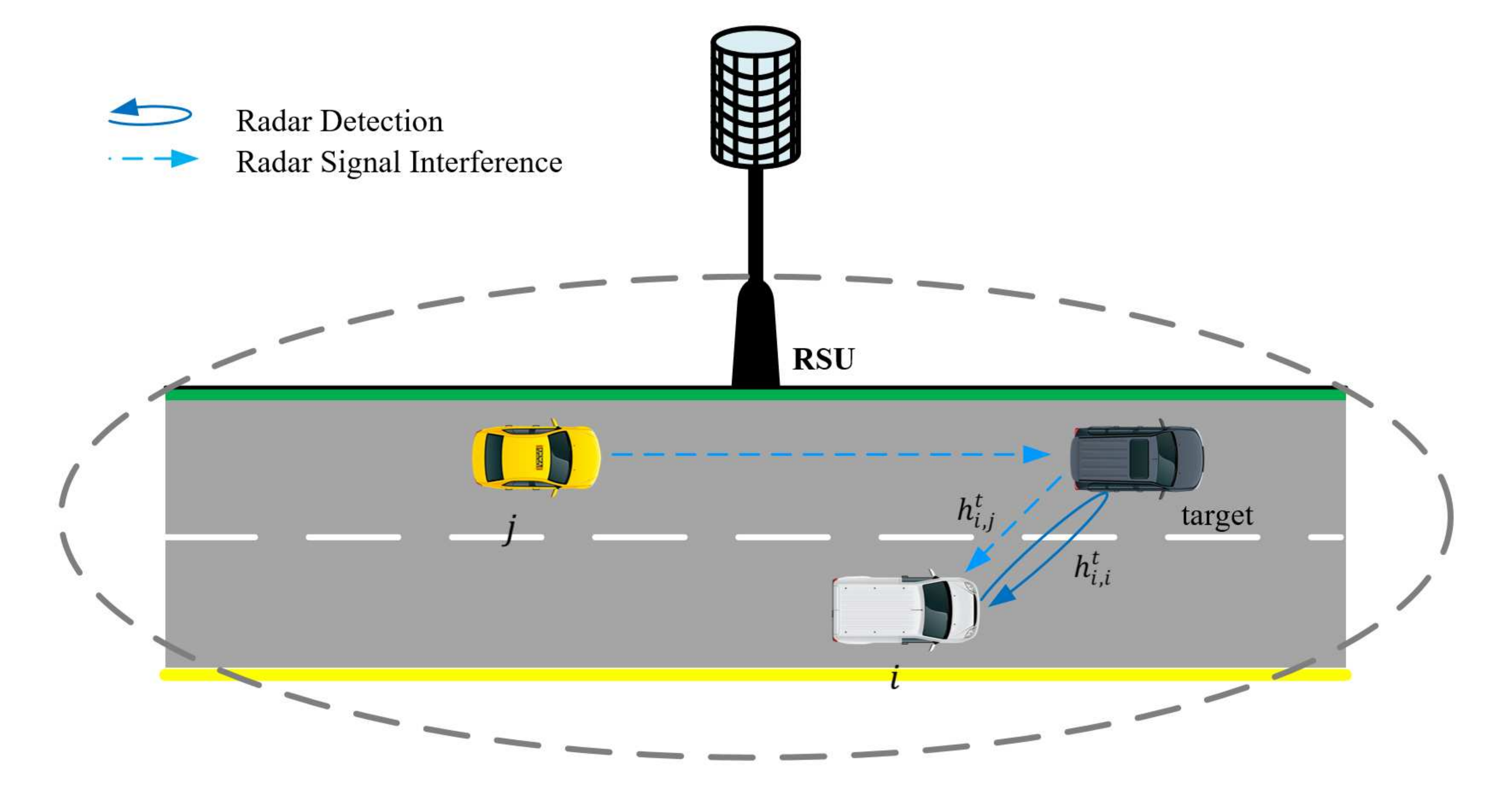}}
\caption{Fig. 4. The propagation gain of radar path in the CAV scenario.}
\label{radar antenna}
\end{figure}

Considering the narrow beam of the mmWave communication system, the main lobe gain is used for the interference analysis. ${G_t}$ and ${G_r}$ represent the antenna transmitting gain and the antenna receiving gain. $\sigma _{i,i}^{RCS}$ represents the Radar Cross Section (RCS) from target to radar $i$, $\sigma _{i,j}^{RCS}$ represents the RCS from radar $j$ to radar $i$, $\lambda $ represents the wave length, ${R_i}$ represents the distance from radar $i$ to target, and ${R_j}$ represents the distance from radar $j$ to target. It is also assumed that all channel gains are fixed during current observation time.

%索引号小
Assuming that vehicle $j$ is the interfering vehicle, when the index of vehicle $j$ is smaller than $n$, the communication interference from vehicle $j$ to vehicle $i$ during the radar sensing duration ${\bar a_n} - {\bar a_{n - 1}}$ of vehicle $i$ is defined by
\begin{equation}\label{radar interference}
n_i^{r - com} = \sum\limits_{j \in {{\cal N}_n}} {{P_j}{G_t}g_{i,j}^{ch - r}{G_r}},
\end{equation}
where ${P_j}$ is the signal transmission power of vehicle $j$, and $g_{i,j}^{ch - r}$ represents the path loss of the communication signal of vehicle $j$ transmitted from vehicle $j$ to vehicle $i$ during the communication duration. ${{\cal N}_n}$ is the vehicle set with the time allocation index number smaller than $n$, and ${\cal N}$ is the set of all vehicles' time indexes.

%索引号大
When the index of vehicle $j$ is bigger than $n$, the radar interference on vehicle $i$ during the radar sensing duration ${\bar a_n} - {\bar a_{n - 1}}$ of vehicle $i$ is
\begin{equation}\label{radar interference}
n_i^{r - rad} = \sum\limits_{j \in {\cal N}\backslash {{\cal N}_n}} {{P_j}h_{i,j}^t} ,
\end{equation}
where ${\cal N}\backslash {{\cal N}_n}$ denotes the vehicle set with the time allocation index bigger than $n$.

%SINR&雷达互信息量
Combining two types of interference above, the SINR of vehicle $i$ in the radar duration ${\bar a_n} - {\bar a_{n - 1}}$ can be expressed as
\begin{equation}\label{radar interference}
\gamma _i^{rad} = \frac{{{P_i}h_{i,i}^t}}{{n_i^{r - rad} + n_i^{r - com} + {N_0}B}},
\end{equation}
where ${N_0}$ and $B$ represent the power spectral density of background noise and the mmWave band bandwidth of JSCIS.

Furthermore, the radar MI for vehicle $i$ in the radar sensing duration can be expressed as
\begin{equation}\label{radarmutualinformation}
\begin{array}{l}
\mathop I_i^{rad}\left( {y;h_{i,i}^t} \right) = H\left( y \right) - H\left( {y|h_{i,i}^t} \right)= \log \left( {1 + \frac{{s{h_{rad}}}}{n}} \right) \\
\;\;\;\;\;\;\;\;\;\;\;\;\;\;\;\;\;\;\;\; = H\left( y \right) - H\left( {n_i^{r - com} + n_i^{r - rad} + n} \right),
\end{array}
\end{equation}
which can be further written as
\begin{equation}\label{radarmutualinformation}
\begin{array}{l}
\mathop I_i^{rad}\left( {y;h_{i,i}^t} \right) = \sum\limits_{n = 1}^{{I_i}} {\left( {{{\bar a}_n} - {{\bar a}_{n - 1}}} \right)\log \left( {1 + \gamma _i^{rad}} \right)}.
\end{array}
\end{equation}

%总的雷达互信息量
Considering the cooperative sensing of $N$ vehicles, the total radar MI is
\begin{equation}\label{radar interference}
{I^{rad}} = \sum\limits_{i = 1}^N {I_i^{rad}}.
\end{equation}

%2)通信传输过程干扰分析
\subsubsection{Signal Analysis During Communication Transmission Process}
\begin{figure}[t]
\centerline{\includegraphics[width=0.5\textwidth]{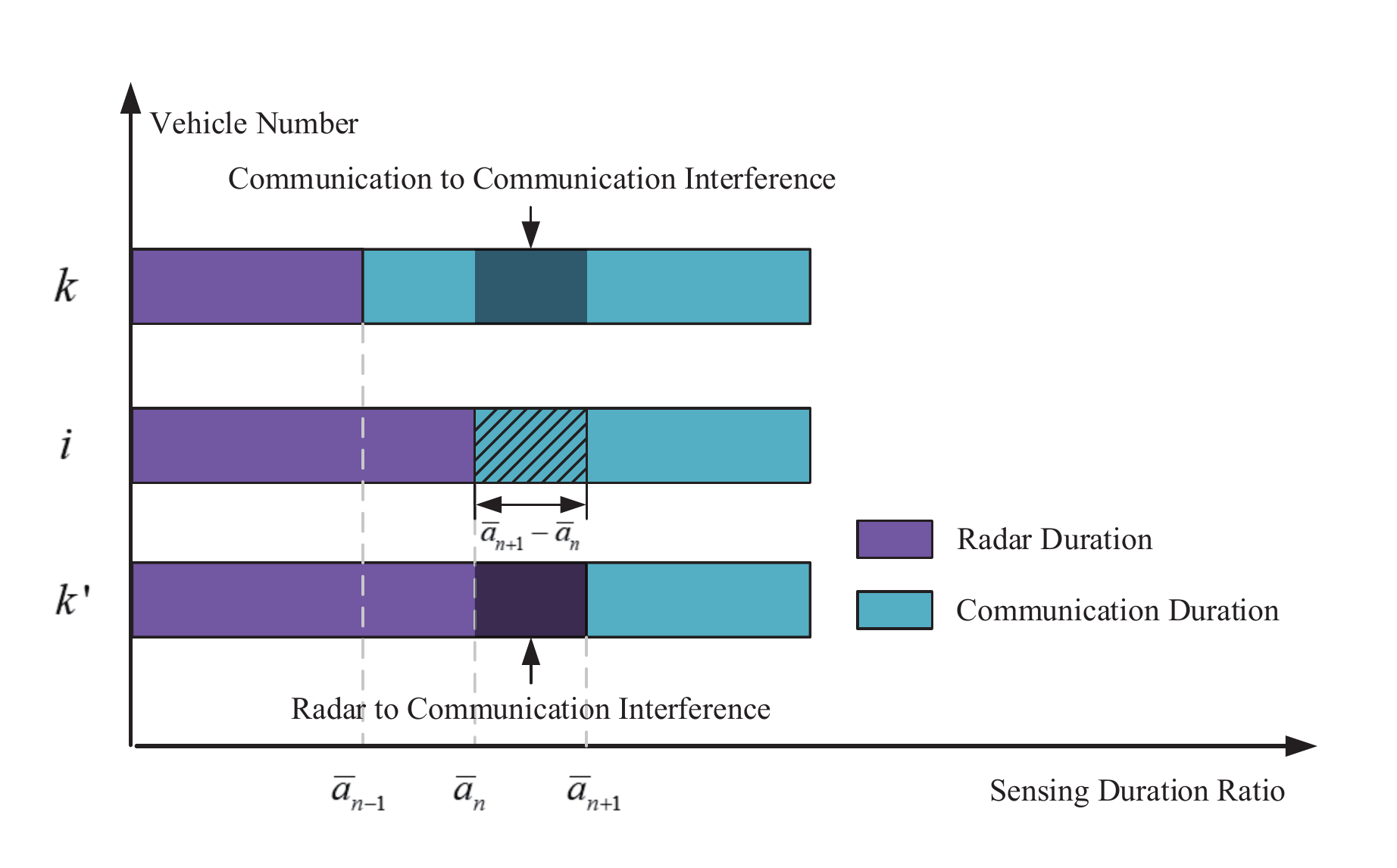}}
\caption{Fig. 5. Interference analysis in the communication duration of JSCIS.}
\label{Interference in communication duration}
\end{figure}

%图的解释
In Fig. 5, for the communication part of vehicle $i$, we assume that vehicle $j$ can communicate with vehicle $i$. In the duration ${\bar a_{n + 1}} - {\bar a_n}$ of vehicle $i$, where ${I_i} \le n \le N$, there are two types of interferences, and the other vehicles are denoted by $k$ and $k'$. Vehicle $i$ will experience the communication signal interference from other vehicles, when the index of other vehicles is smaller than $n$. When the index of other vehicles is bigger than $n$, vehicle $i$ will experience the radar signal interference from other vehicles. Thus, the received communication signal is depicted by
\begin{equation}\label{communicationsignal}
{y_{com}} = {s_{com}}{h_{com}} + {n^{c - com}} + {n^{c - rad}} + n,
\end{equation}
where ${n^{c - com}}$ is the interference caused by communication signals of other vehicles to communication signals of vehicle $i$, and ${n^{c - rad}}$ is the interference caused by radar signals of other vehicles to communication signals of vehicle $i$.

%索引号小
Assuming that vehicle $k$ is the interfering vehicle, when the time allocation index of vehicle $k$ is smaller than $n$, the communication interference of vehicle $k$ to the current vehicle $i$ in the communication duration can be expressed as
\begin{equation}\label{radar interference}
n_i^{c - com} = \sum\limits_{k \in {{\cal N}_n},k \ne j} {{P_k}{G_t}g_{i,k}^{ch}{G_r}},
\end{equation}
where ${P_k}$ is the signal transmission power of vehicle $k$, and $g_{i,k}^{ch}$ represents the path loss from vehicle $k$ to vehicle $i$.

%索引号大
When the index of the vehicle $k$ is bigger than $n$, the radar interference from vehicle $k$ in the radar duration to vehicle $i$ in the communication duration is
\begin{equation}\label{radar interference}
n_i^{c - rad} = {\sum _{k \in {\cal N}\backslash {{\cal N}_n},k \ne j}}{P_k}\frac{{{G_t}{G_r}{\lambda ^2}}}{{{{(4\pi )}^2}d_{i,k}^2}},
\end{equation}
where ${d_{i,k}}$ represents the distance between vehicle $i$ and vehicle $k$.

%SINR
Combining these two types of interferences, the SINR of vehicle $i$ in the communication duration ${\bar a_{n + 1}} - {\bar a_n}$ can be expressed as
\begin{equation}\label{radar interference}
\gamma _i^{com} = \frac{{{P_i}{G_t}g_{i,j}^{ch}{G_r}}}{{n_i^{c - com} + n_i^{c - rad} + {N_0}B}}.
\end{equation}

%通信信息率
Then, the communication rate of vehicle $i$ is defined by
\begin{equation}\label{radar interference}
R_i^{com} = B\sum\limits_{n = {I_i}}^N {({{\bar a}_{n + 1}} - {{\bar a}_n}){{\log }_2}(1 + \gamma _i^{com})}.
\end{equation}

%通信信息量
Therefore, the communication channel capacity can be expressed as
\begin{equation}\label{radar interference}
I_i^{com} = \left( {1 - {a_{{I_i}}}} \right)T \cdot B\sum\limits_{n = {I_i}}^N {\left( {{{\bar a}_{n + 1}} - {{\bar a}_n}} \right){{\log }_2}\left( {1 + \gamma _i^{com}} \right)}.
\end{equation}

Considering the cooperative sensing of $N$ vehicles, the total communication channel capacity is
\begin{equation}\label{radar interference}
{I^{com}} = \sum\limits_{i = 1}^N {I_i^{com}}.
\end{equation}

%%3)AOI相关分析
\subsubsection{AoI Analysis}
\begin{figure}
\centering
\includegraphics[angle=0,width=0.5\textwidth]{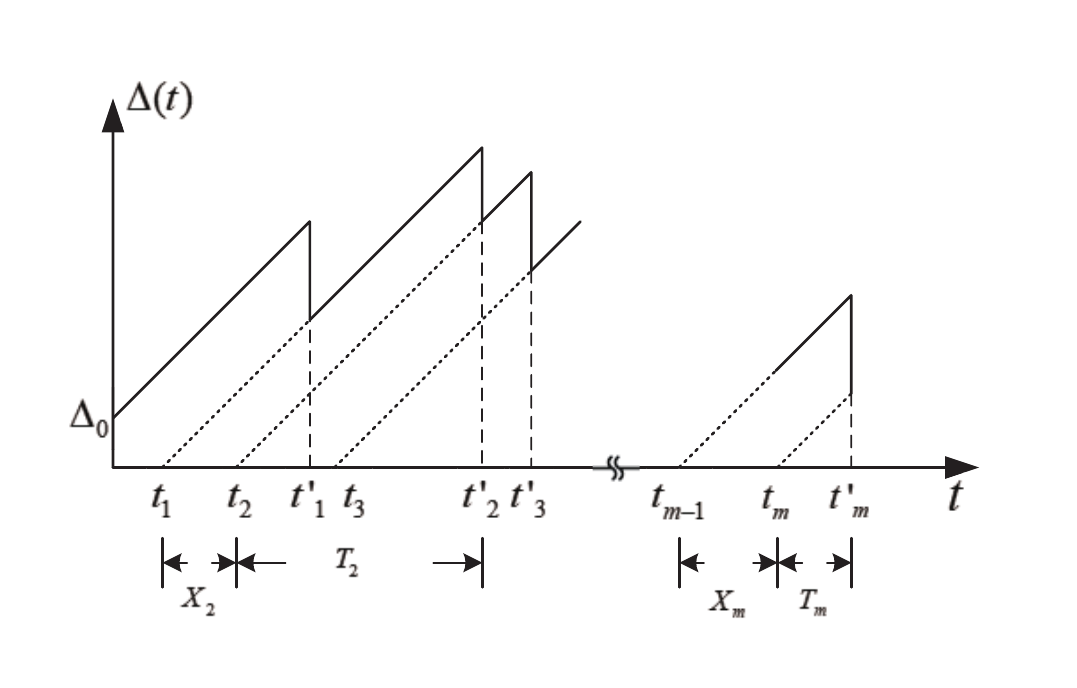}
\caption{Fig. 6. Status update age for a system with a FCFS queue.}
\label{AoI}
\end{figure}

AoI stands for the timeliness of information, which should be as small as possible to meet the service requirements [18]. In this paper, to evaluate the timeliness of raw sensing data sharing among vehicles and RSUs, we use AoI to model the performance of $M/M/1$ queuing problem to achieve the best time duration allocation ratio for sensing and communication dual functions for each vehicle in the proposed CAV scenario. Furthermore, we analyze the effects of different duration allocation ratios on the packet loss probability and the average AoI.

Fig. 6 shows a sample variation of AoI $\Delta \left( t \right)$ as a function of time $t$. The first status update is generated at ${t_1}$, followed by updates at ${t_2},{t_3}, \ldots ,{t_m}$. AoI at time $t$ is defined as $\Delta \left( t \right) = t - u\left( t \right)$, where the $u\left( t \right)$ is the timestamp of the most recent information at the receiver. $X$ is the interval between two arrived updates, and $T$ is the time of one update served in the system. Both $X$ and $T$ are random variables. The average AoI can be calculated during the observation interval $\left( {0,\tau } \right)$ as [18]

\begin{equation}\label{averageAOI1}
{\left\langle {{\Delta _i}} \right\rangle _\tau } = \mathop {\lim }\limits_{\tau  \to \infty } \frac{1}{\tau }\int_0^\tau  {\Delta \left( t \right)dt},
\end{equation}
which can be further converted to
\begin{equation}\label{averageAOI1}
{\Delta _i} = \Lambda \left( {\frac{1}{2}E\left[ {{X^2}} \right] + E\left[ {XT} \right]} \right).
\end{equation}

\begin{figure}
\centering
\includegraphics[angle=0,width=0.5\textwidth]{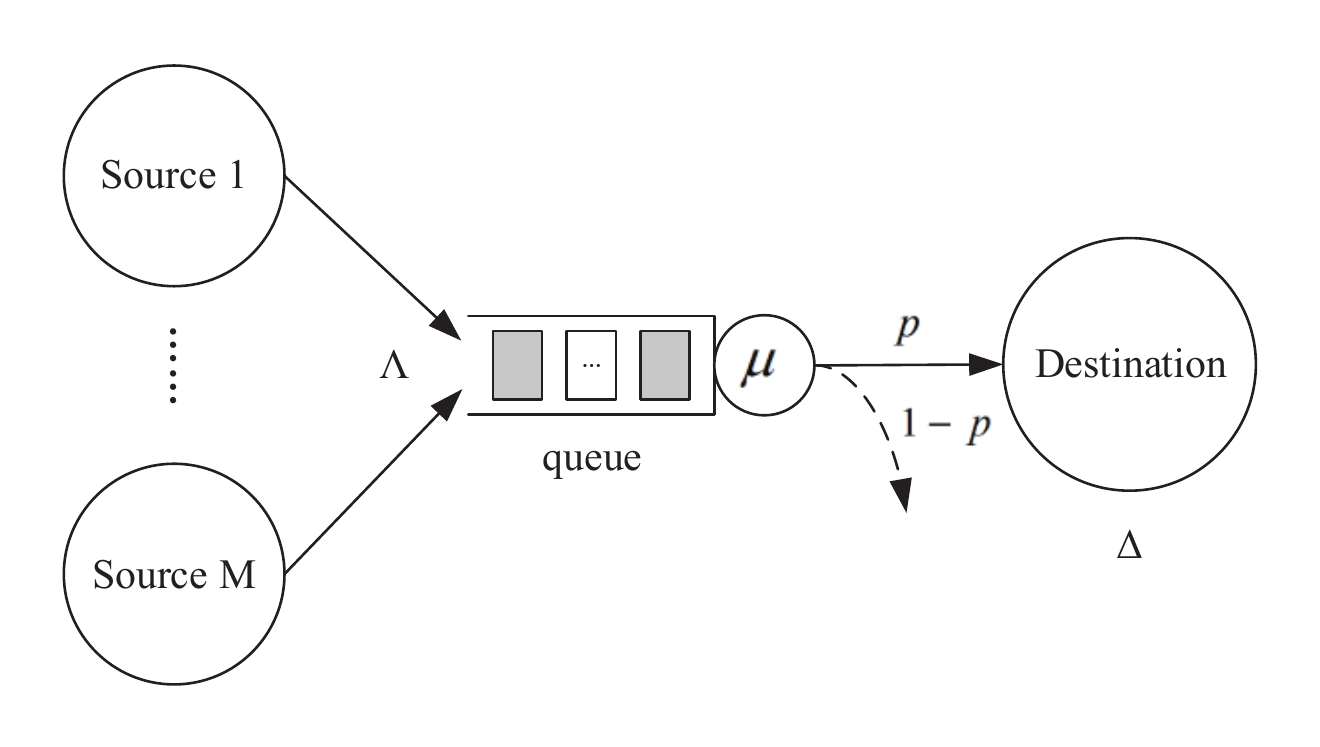}
\caption{Fig. 7. The FCFS queue model of multiple sources to one destination with packet loss.}
\label{queue}
\end{figure}

The information system in this paper is assumed as a centralized network of multi-source status update with packet loss, in which all vehicles transmit their detection information via the communication links to the central vehicle. As shown in Fig. 7, the FCFS queue model of multiple sources $i = 1,2, \ldots ,N$ with packet loss is depicted. The FCFS scheduling strategy can effectively ensue the timeliness of the radar detection information transmission. Considering the timeliness requirement of information transmission between vehicles as well as the cache and computation capacities of vehicles, the computation model is modeled as a typical M/M/1 queue based on [30], and each vehicle only carries out serial calculation for task packets. The sources generate data packets according to a Poisson process with an arrival rate $\Lambda$. The service time for each packet is exponentially distributed with the service rate $\mu$. The server load rate is $\rho {\rm{ = }}\Lambda {\rm{/}}\mu$. In [31], the average AoI for the $M/M/1$ queue is given by
\begin{equation}\label{average-MM1}
{\Delta _i} = \frac{1}{\mu }\left( {1 + \frac{1}{\rho } + \frac{{{\rho ^2}}}{{1 - \rho }}} \right).
\end{equation}

In the communication system, the packet is successfully transmitted only if the SINR at the receiver does not fall below a threshold ${\gamma _c}$. Combined with the interference analysis in Section \uppercase\expandafter{\romannumeral2}-B, the success transmission probability $p$ can be expressed as

\begin{equation}\label{comsuccess}
p = {\rm P}\left\{ {\left( {\mathop {\min }\limits_{i \in \left\{ {1,2, \ldots ,N} \right\}} \gamma _i^{com}} \right) > {\gamma _c}} \right\}.
\end{equation}

Based on [32], (23) can be simplified as

\begin{equation}\label{Simp_comsuccess}
p = \exp \left( { - {{\left( {\frac{\varphi }{{2\pi }}} \right)}^2}a\lambda \pi {{\left( {\frac{{{\gamma _c}}}{{{d^{ - \alpha }}}}} \right)}^{\frac{{\rm{2}}}{\alpha }}}} \right),
\end{equation}
where $\varphi$ is the width of the main lobe of antenna, which is fixed to $\pi /6$, $a$ is the time allocation ratio, $\lambda$ is the network density. We assume that vehicles follow a homogeneous Poisson point process with the intensity $\lambda$. $d$ is the distance between vehicles, which is limited to 200 m. And the packet loss rate is $1-p$.

Considering the packet loss probability, the server load rate $\rho$ turns to $\rho ' = p\frac{\Lambda }{\mu }$, and the average AoI of $M/M/1$ queue (22) turns to

\begin{equation}\label{averageloss}
{\Delta _i} = \frac{1}{\mu }\left( {1 + \frac{1}{{\rho '}} + \frac{{\rho {'^2}}}{{1 - \rho '}}} \right).
\end{equation}

%4)目标函数
\subsubsection{Objective Function}
According to the existing research, the using of current mmWave on-board radars in the middle and high frequency bands can obtain the location information of vehicles accurately, such as distance and direction, which can be used to establish the vehicle network topology via the communication links among vehicles. In [33], a new millimeter-wave distance-measurement sensor prototype was developed for the first time, and small errors between measured and actual distances were achieved. The mmWave radar can obtain the network topology through detecting with low transmission power and small absolute error, then the detection information can be transmitted to the central vehicle via communication links. The optimized beam can be determined by the beam training algorithm [34]. The performance optimization of JSCIS is a multi-objective (dual-objective) problem that optimizes both radar and communication systems's performance at the same time. The solution to this problem is to obtain the Pareto optimality for the objective function.

To maximize the amount of information detected by radars, the performance optimization of JSCIS can be modeled as a maximization problem of radar total MI with the communication channel capacity as a constraint. Therefore, the optimization problem can be defined as
\begin{equation}\label{radar interference}
%\;是空格的代码
\begin{array}{l}
\mathop {\max }\limits_a \;{I_{rad}} \\
s.t.\;\;\;{\rm{C}}1:{a_n} \in {\Omega _t},\;\forall n,\\
\;\;\;\;\;\;\;\;{\rm{C}}2:I_i^{rad} \le I_i^{com},\\
\;\;\;\;\;\;\;\;{\rm{C}}3:{\Delta _i} \le \Delta _i^{\max },\\
%\;\;\;\;\;\;\;\;{\rm{C3}}:\min {\Delta _i},\left( {\mathop {\min }\limits_{i \in \left\{ {1,2, \ldots ,N} \right\}} \gamma _i^{com}} \right) > {\gamma _c},
\end{array}
\end{equation}
where constraint C1 provides the available radar duration set for each vehicle, constraint C2 is to guarantee that the radar information of each vehicle can be efficiently transmitted within the communication time of $T$, constraint C3 is used to ensure that the data transmission efficiency is below a certain threshold in case of a success packet transmission.

%第三部分：基于非合作博弈的时间资源分配方法
\section{Time Resource Allocation Method Based on Non-cooperative Game}
%%第三章的总领性段落。
The resource allocation optimization problem for multiple CAVs is formulated as a non-cooperative game, and the feasibility and existence of the pure strategy NE are proved. To achieve the NE of the non-cooperative game problem, we proposed the CTRA algorithm, and the convergence of the CTRA algorithm is also proved theoretically.

%A.非合作博弈问题的建立
\subsection{The Formulation of the Non-cooperative Game}
It can be observed that the optimization problem of JSCIS is combinatorial and non-convex. The combinational nature comes from the constraint C1. Based on the constraint C2 in (26), the problem is non-convex. Although the exhaustive method can obtain the optimal time resource allocation strategy, it is computationally infeasible for CAVs and also not effective for the system design. In order to reduce the computational complexity and select a suitable resource allocation strategy, we propose a utility function and use the non-cooperative game method to solve this problem.

%%A.纳什均衡分析
%\subsection{Nash Equilibrium Analysis}

The duration occupied by radar process will affect the time allocation and the information transmission in the communication process, and vice versa. Therefore, the optimal time allocation problem of JSCIS is defined as a non-cooperative game problem as
\begin{equation}\label{radar interference}
{\cal G} = \left\langle {{\cal K},{{\{ {{\cal A}_m}\} }_{m \in {\cal K}}},{{\{ {U_m}\} }_{m \in {\cal K}}}} \right\rangle,
\end{equation}
where ${\cal K} = {\rm{\{ 1,2,}} \cdots ,N{\rm{\} }}$ is the set of players (i.e., vehicles in the CAVs scenario). ${{\cal A}_m}$ is the set of available pure strategies for all players $m$, and ${U_m} = U$ is the utility function of player $m$.

%效用函数
The goal of each player is to maximize its own utility function by choosing a suitable time allocation strategy. Considering the objective function and constraints, the utility function can be defined as
\begin{equation}\label{radar interference}
U = {I^{rad}} + \eta \sum\limits_{i = 1}^N {\Phi (I_i^{com},\;I_i^{rad})},
\end{equation}
where $\eta $ is the penalty scalar whose unit is ``bps", and $\Phi (x,y)$ is the penalty function, which satisfies that $\Phi (x,y) =  - 1$, if $x < y$, and $\Phi (x,y) =  0$, if $x \ge y$. It is worth noting that the first term in (28) corresponds to the radar total MI, and the second term in (28) indicates the limitation of the communication channel capacity. If the communication channel capacity is smaller than the radar MI, the utility function will be punished.

Under the condition that $\eta  > \max {I^{rad}}$ and each player has a feasible solution that satisfies the constraints C1 and C2, there will be $U < 0$ if an arbitrary player chooses a strategy that violates the constraint C2. It means that any strategy profiles violating the constraint C2 will never be the optimal solution. Therefore, the optimal solution for maximizing $U$ is equivalent to the optimal solution of the objective function in (26).

Let ${A_m} \in {{\cal A}_m}$ denotes the strategy of player $m$, and ${A_{ - m}} \in {{\cal A}_1} \times  \cdots  \times {{\cal A}_m} \times  \cdots {{\cal A}_N}$ represents the strategy profiles of all players excluding player $m$, where $\times$ is the Cartesian product. In this paper, ${A_m} = {a_m}$ represents the radar duration proportion of vehicle $m$ in each frame. Given a strategy of an arbitrary player $m$, ${A_m} \in {\cal A}{}_m$, and an alternative strategy ${A'_m} \in {{\cal A}_m}$, while the strategies of other players remain unchanged, then we have
\begin{align}
\begin{array}{l}
\;\;\;\;{U_m}({A_m},{A_{ - m}}) - {U_m}({A'_m},{A_{ - m}})\\
= U({A_m},{A_{ - m}}) - U({A'_m},{A_{ - m}}).
\end{array}
\end{align}

The formulated game ${\cal G}$ is a potential game with the potential function $U$. The selection of each player's strategy can be mapped into the potential function. According to the properties of potential games, each potential game has at least one pure strategy Nash equilibrium (NE). If there is no constraint C2, each NE of game ${\cal G}$ will be the solution. However, due to the constraint C2, whether a pure strategy NE exists or not is unclear. Therefore, the existence of feasible NE for the proposed game ${\cal G}$ need to be proved. To simplify the subsequent proof and derivation process, the maximum radar MI is defined as $\tilde \eta  = \max {I^{rad}}$.

%B.纳什均衡解的可行性
\subsection{The Feasibility of Pure Strategy Nash Equilibrium}
Assuming that there are enough time slots for each scheduling cycle, we analyze the feasibility of pure strategy NE for the proposed game.

%定义1（纯策略NE）的定义
\textbf{\emph{Definition 1 (Pure Strategy NE):}} The strategy profile $\left( {A_1^*, \ldots ,A_N^{\rm{*}}} \right)$ is a pure strategy NE of ${\cal G}$. If any $m \in {\cal K}$ has an alternate ${A_m} \ne A_m^*$, the following conditions will be always true for all ${A_m} \in {{\cal A}_m}$ [35],
\begin{equation}\label{pure}
{U_m}(A_m^*,A_{ - m}^*) \ge {U_m}({A_m},A_{ - m}^*).
\end{equation}

\newtheorem{thm1}{Theorem}
\begin{thm}\label{thm1}
If $\eta  \ge \tilde \eta $ , each scheduling cycle consists of enough time slots, and there is no impact on other vehicles when each participating vehicle adjusts its time allocation under the constraint C2. Then the pure strategy NE of the proposed game ${\cal G}$ must be feasible.

\begin{proof}
Assume that the strategy profile $\left( {A_1^*, \ldots ,A_N^*} \right)$, which violates the constraint C2, is a pure strategy NE of ${\cal G}$. Then, there must be some vehicles whose communication channel capacity is smaller than radar MI. We denote these vehicles by player $m'\left( {1 \le m' \le N} \right)$. Since we assume that each scheduling cycle consists of enough time slots, player $m'$ must be able to choose another strategy ${A'_{m'}}$ with a smaller radar detection duration to make its radar MI be smaller than the communication data information. Then, we have
\begin{align}
\begin{array}{l}
{U_{m'}}(A_{m'}^*,A_{ - m'}^*) - {U_{m'}}({{A'}_{m'}},A_{ - m'}^*)\\
\;\;\;\;\;\; = {I^{rad}}(A_{m'}^*,A_{ - m'}^*) - {I^{rad}}({{A'}_{m'}},A_{ - m'}^*)\\
\;\;\;\;\;\;\;\;\;\; + \eta [\sum\limits_{i = 1}^N {\Phi (I_i^{com}(A_{m'}^*,A_{ - m'}^*),\;I_i^{rad}(A_{m'}^*,A_{ - m'}^*))} \\
\;\;\;\;\;\;\;\;\;\;- \sum\limits_{i = 1}^N {\Phi (I_i^{com}({{A'}_{m'}},A_{ - m'}^*),\;I_i^{rad}({{A'}_{m'}},A_{ - m'}^*))} ]\\
\;\;\;\;\;\; = {I^{rad}}(A_{m'}^*,A_{ - m'}^*) - {I^{rad}}({{A'}_{m'}},A_{ - m'}^*) - \eta  < 0.
\end{array}
\end{align}

Apparently, (31) contradicts the assumption that $(A_1^*, \cdots ,A_N^*)$ is a pure strategy NE of ${\cal G}$. Therefore, the pure strategy NE of ${\cal G}$ must be feasible if the conditions of this theorem hold, which concludes this proof.
%证明结束

\end{proof}
\end{thm}

According to \textbf{Theorem 1}, the solution of game ${\cal G}$ that meets certain conditions is feasible, which means that when certain conditions are not met, not all pure strategies NE can work. Therefore, the existence of pure strategies NE of ${\cal G}$ is described below.
%定理1表明了在持续时间互不影响和时隙数足够多的情况下，保证了所提博弈解决方案的可行性。
%接下来调查博弈G可行纯策略纳什均衡的存在性。

%C.纳什均衡解的存在性
\subsection{The Existence of pure strategy Nash Equilibrium}

%定理2
%η，足够的时隙，则给出的博弈至少有一个可行纯策略纳什均衡，并且问题P的最优解决方案组成了博弈G 的纯策略NE
\newtheorem{thm2}{Theorem}
\begin{thm}\label{thm1}
If $\eta  \ge \tilde \eta $ and each scheduling cycle consists of enough time slots, the proposed game ${\cal G}$ has at least one feasible pure strategy NE.

\begin{proof}
%证明
Following the proof in [36], we assume that $\left( {A_1^*, \ldots ,A_N^*} \right)$ can maximize $U$. If these conditions hold, then we have $U\left( {A_1^*, \ldots ,A_N^*} \right) > 0$. If $\left( {A_1^*, \ldots ,A_N^*} \right)$ is infeasible, we have $U\left( {A_1^*, \ldots ,A_N^*} \right) < 0$, which contradicts the above inequality. Hence, $\left( {A_1^*, \ldots ,A_N^*} \right)$ must be feasible, and we have $U(A_1^*, \cdots ,A_N^*) = {R^{rad}}(A_1^*, \cdots ,A_N^*)$, when $\eta  \ge \tilde \eta $ holds. Since $U(A_1^*, \cdots ,A_N^*)$ satisfy
\begin{align}
\begin{array}{l}
\;\;\;\;U(A_1^*, \cdots ,A_N^*)\\
\ge U({A_1}, \cdots ,{A_N}),\forall ({A_1}, \cdots ,{A_N}) \in {\{ {{\cal A}_m}\} _{m \in {\cal K}}},
\end{array}
\end{align}
then we have ${R^{rad}}(A_1^*, \cdots ,A_N^*) \ge {R^{rad}}({A_1}, \cdots ,{A_N})$ for any feasible strategy profile $\left( {{A_1}, \ldots ,{A_N}} \right)$ with the conditions of \textbf{Theorem 2}. Therefore, $\left( {A_1^*, \ldots ,A_N^*} \right)$ is the optimal solution to the objective function. Obviously, there is no other feasible strategy profile that can further improve the utility function, which means that no strategy profiles $({A_1}, \cdots ,{A_N}) \in {\{ {{\cal A}_m}\} _{m \in {\cal K}}}$ satisfies
\begin{align}
\begin{array}{l}
\;\;\;\;{U_m}({A_1}, \cdots ,{A_N}) = U({A_1}, \cdots ,{A_N})\\
> {U_m}(A_1^*, \cdots ,A_N^*) = U(A_1^*, \cdots ,A_N^*),\;\forall m \in {\cal K}.
\end{array}
\end{align}

Suppose for $\forall m \in {\cal K}$, if ${A_m} \in {{\cal A}_m}$ and ${A_m} \ne A_m^*$ is other strategies of player $m$, then
\begin{align}
{U_m}({A_m},A_{ - m}^*) \le {U_m}(A_m^*,A_{ - m}^*).
\end{align}

Similarly, no players can unilaterally change their strategies to improve the utility. Therefore, according to the definition of NE, $(A_1^*, \cdots ,A_N^*)$ is the pure strategy NE of ${\cal G}$, which concludes this proof.

\end{proof}
\end{thm}

\textbf{Theorem 2} denotes that there is at least one feasible pure strategy NE. If there is a single pure strategy NE, the only pure strategy NE must be the Pareto optimal solution when the condition of \textbf{Theorem 2} holds.

%D.基于非合作博弈的集中式时间资源分配算法
\subsection{CTRA Algorithm Design Based on Non-cooperative Game}
According to the above analysis, based on the best strategy game, a low-complexity CTRA algorithm is designed to achieve the pure strategy NE.
%算法复杂度分析
Assuming that $m$ represents the total number of iterations, $n$ represents the number of vehicles, and $p$ represents the number of sub-frames contained in each frame, the algorithm needs to calculate $n \times p$ times of utility function for each iteration. Therefore, the algorithm complexity can be expressed by ${\cal O}\left( {mnp} \right)$.

\begin{algorithm}[t]
\caption{Centralized Time Resource Allocation (CTRA) Algorithm} \label{alg_1}
\begin{algorithmic}[1]
\STATE \textbf{Initialization}: Initialization strategy ${A_m} = {\Omega _t}(1)$, $1 \le m \le N$, iteration number $t=0$, $U_{\max}=0$;
\STATE \textbf{repeat}
\STATE \quad \textbf{for} $m = 1:N$
\STATE \quad \quad $A_m^{t + 1} = \mathop {{\mathop{\rm argmax}\nolimits} }\limits_{{A_m} \in {{\cal A}_m}} {U_m}({A_m},{A_{ - m}})$;
\STATE \quad \quad if $U>U_{\max}$
\STATE \quad \quad update $A_m^t{\rm{ = }}A_m^{t + 1}$, $U_{\max}=U$;
\STATE \quad \textbf{end for}
\STATE \quad update $t = t + 1$.
\STATE \textbf{until} ${U_m}(A_m^t,A_{ - m}^t) = {U_m}(A_m^{t - 1},A_{ - m}^{t - 1})$, $\forall m \in {\cal K}$.
\STATE \textbf{return} $({A_m},{A_{ - m}})$.
\end{algorithmic}
\end{algorithm}
Since not all pure policy NEs of game ${\cal G}$ are feasible, it is necessary to start \textbf{Algorithm 1} from a feasible strategy profile to ensure that it can converge to a feasible pure policy NE. Therefore, the initial detection duration for each vehicle is set to be the minimum in its available duration set, which ensures that the initial strategy profile is feasible under previous assumption. Based on the proof in [36], the convergence of \textbf{Algorithm 1} can be proved.

%定理3
\newtheorem{thm3}{Theorem}
\begin{thm}\label{thm1}
If $\eta  \ge \tilde \eta $ and each scheduling cycle consists of enough time slots, the proposed \textbf{Algorithm 1} converges to a feasible pure strategy NE of ${\cal G}$ in finite steps from any initial feasible strategy profiles.

%证明
\begin{proof} Starting from any initial feasible strategy profiles, according to the best response procedure, we have
\begin{equation}\label{aaa}
{U_m}(A_m^{t + 1},A_{ - m}^t) - {U_m}(A_m^t,A_{ - m}^t) > 0,\forall m \in {\cal K},
\end{equation}
which states that the utilities of all players are strictly increasing in each iteration of CTRA. Suppose that ${U^*}$ is the maximum value of $U$, we have ${U^*} < \infty $, since the number of each player's strategies and the number of players are both finite. Moreover, the proposed game ${\cal G}$ is a potential game with the potential function $U$, then we have
\begin{align}
\begin{array}{l}
\;\;\;\;{U_m}(A_m^{t + 1},A_{ - m}^t) - {U_m}(A_m^t,A_{ - m}^t)\\
= U(A_m^{t + 1},A_{ - m}^t) - U(A_m^t,A_{ - m}^t),\forall m \in {\cal K}.
\end{array}
\end{align}

According to (35) and (36), we can conclude that each update of a player's strategy at each iteration will result in a strictly increasing quantity of $U$. Since ${U^*} < \infty $, then there must exist a ${t^*}\left( {0 \le {t^*} < \infty } \right)$ such that ${U_m}(A_m^{{t^{\rm{*}}}},A_{ - m}^{{t^*}}) = {U^*}$, when ${t^*}$ is sufficiently large. That is, the strictly increasing procedure of the proposed CTRA algorithm converges within finite steps, which concludes this proof.
\end{proof}
\end{thm}

%第四部分：仿真与结果分析
\section{Results and Analysis}
Both software simulation and hardware testbed are designed and developed to verify the feasibility of the proposed JSCIS and the performance of the CTRA algorithm. The effect of different time duration ratios on packet loss probability and average AoI is simulated and analyzed. Besides, by comparing the performance of different algorithms, the solution of the time resource allocation optimization problem is verified and the performance of the proposed algorithm under different impact factors is evaluated with discussions in detail. Finally, the hardware testbed is setup and results are analyzed to verify the feasibility of the proposed JSCIS.

%A.软件仿真的结果分析
\subsection{Software Simulation Setup and Results Analysis}
%软件仿真的参数设置说明
The key simulation parameters are listed in Table II. The antenna gains of transmitter and receiver of radar system are the same as those of communication system, which are set to 18 dB inside the main lobe beamwidth of $\varphi$. The transmit power of communication and radar systems is set to 10 W [37]. According to the link budget analysis, the maximum distance between vehicles is set to 200 m to ensure the effective radar detection and communication. The number of subframes ${N_s}$ in each frame is set to 14 based on [38]. The JSCIS operates at 28 GHz, and the bandwidth is 800 MHz [38]-[39]. The normalization parameter $\tilde \eta $ in the CTRA algorithm is 10 Gbps. We assume that the RCS model of radar system has a uniform reflectivity, which is ${\sigma ^{RCS}} = 1$.

\begin{table}[t]
\centering
 \caption{ \label{parameter setting}Table II. Key simulation parameters of JSCIS.}
 \begin{center}
 \begin{tabular}{l l l}
 \hline
 \hline
  {Parameter} & {Value}\\
  \hline
  {Carrier frequency} & {28 GHz}\\
  {System bandwidth $B$} & {800 MHz}\\
  {Thermal noise power spectral density ${N_{\rm{0}}}$} & {-174 dBm/Hz}\\
  {Shadow fading standard deviation} & {8 dB}\\
  {Transmit power} & {10 W}\\
  {Tx/Rx antenna gain} & {18 dB}\\
  {Communication SINR threshold ${\gamma _c}$} & {5}\\
  {Distance between vehicles} & {200 m}\\
  {Normalization parameter $\tilde \eta $} & {10 Gbps}\\
  {Maximum acceptable average AoI $\Delta _i^{\max }$} & {4}\\

  \hline
  \hline
 \end{tabular}
 \end{center}
\end{table}

%关于AoI部分的仿真结果及分析
\begin{figure}[t]
\centerline{\includegraphics[width=0.5\textwidth]{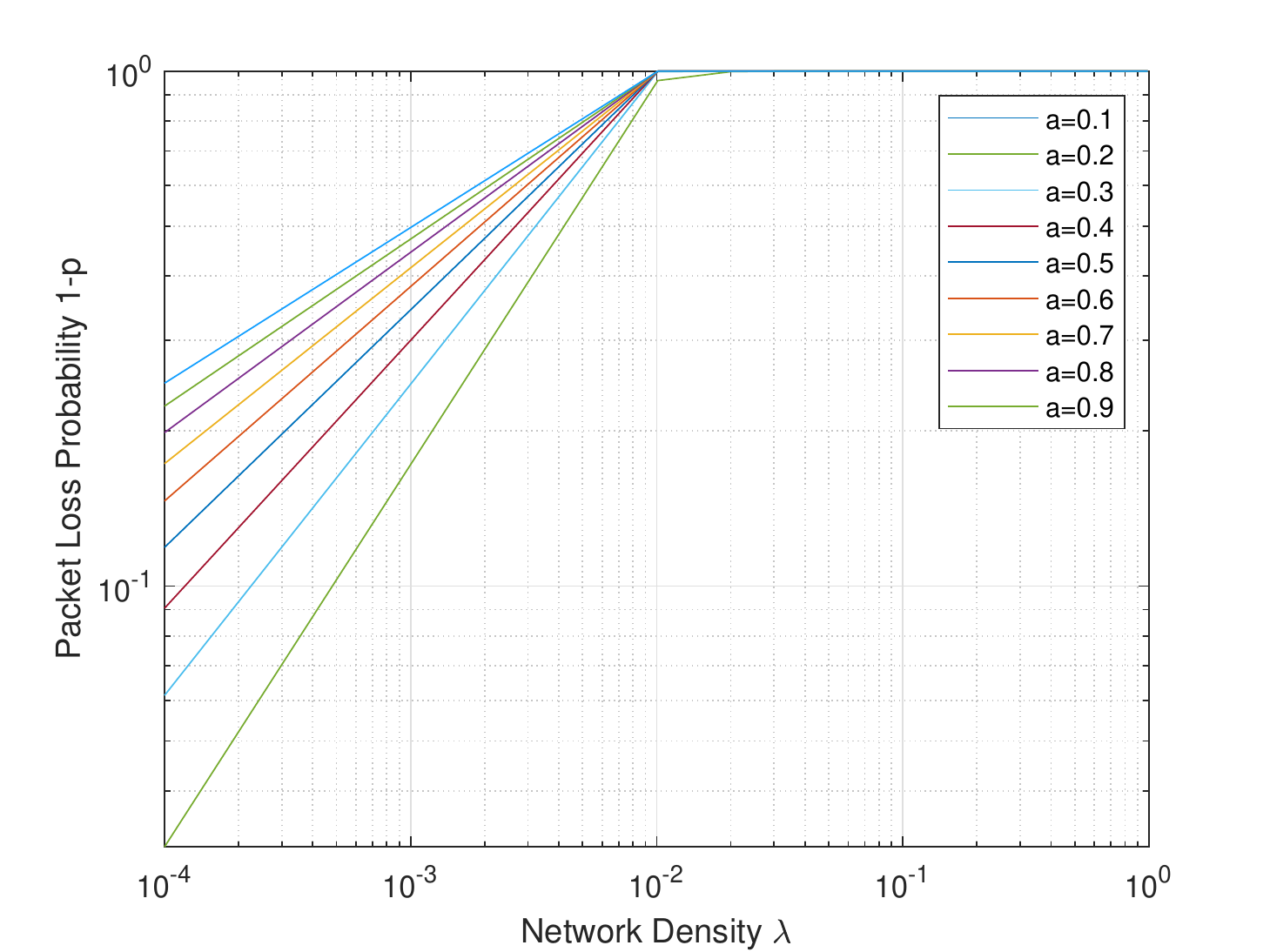}}
\caption{Fig. 8. The change of packet loss rate under different time allocation ratios.}
\label{packloss}
\end{figure}

The influence of different time allocation ratio and network density on the packet loss rate are shown in Fig. 8. And the number of vehicles is set to 10. As the network density increases, the packet loss rate gradually increases, and the performance of communication system decreases. When $\lambda  > {10^{ - 2}}$, the performance of communication system deteriorates, and the packet loss rate is approaching 1. Moreover, when the network density is constant, with the increase of time allocation ratio, the packet loss rate also increases. When $a=0.1$, the communication performance is the best.

\begin{figure}[t]
\centerline{\includegraphics[width=0.5\textwidth]{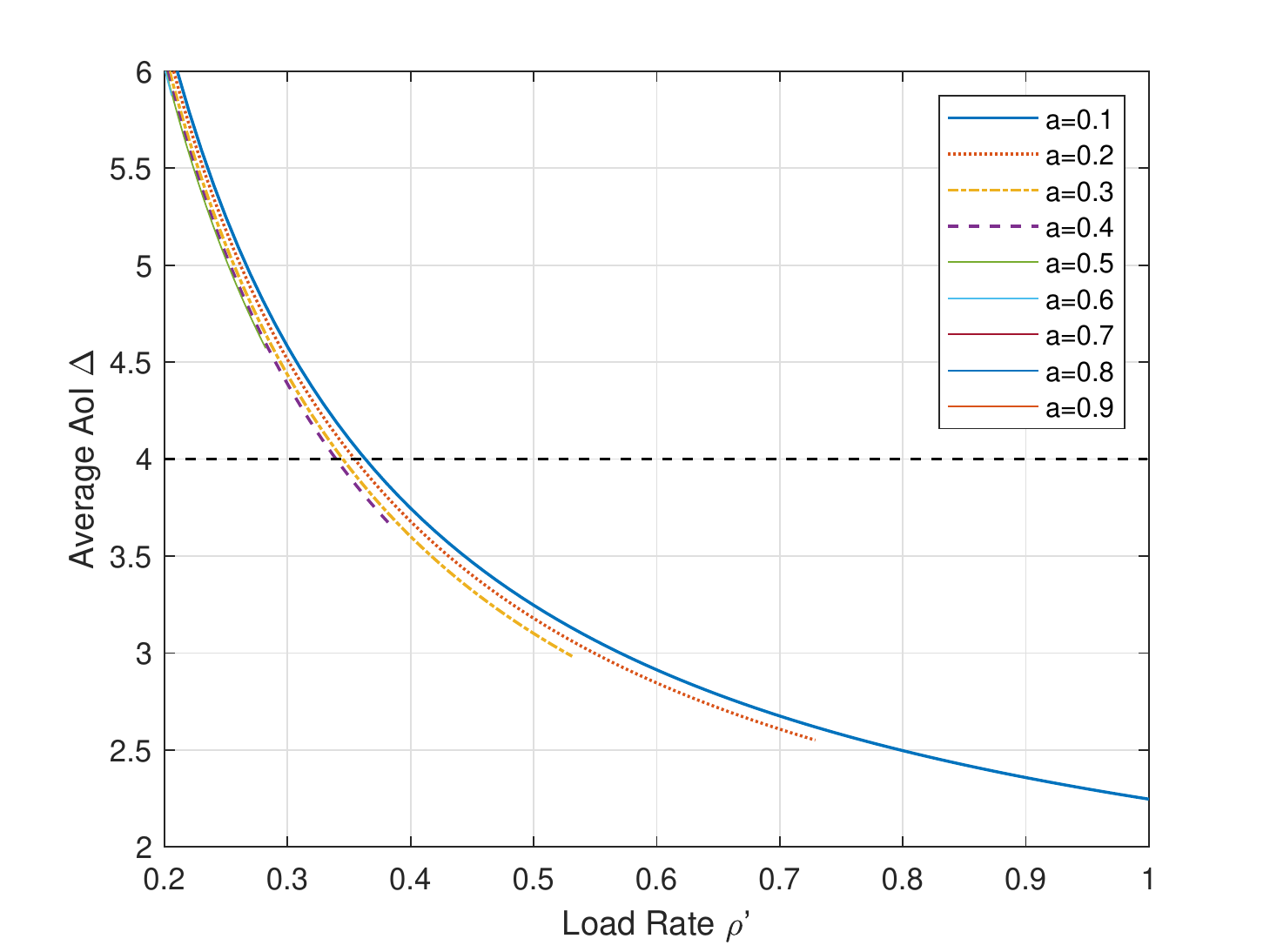}}
\caption{Fig. 9. The change of average AoI under different time allocation ratios.}
\label{AverageAge}
\end{figure}

The relationship among average AoI $\Delta$, time ratio $a$ and server load rate $\rho '$ is shown in Fig. 9. With the increase of allocation ratio, the average AoI decreases. When $a={0.1,0.2,0.3,0.4}$, the average AoI can satisfy the limit of the maximum average AoI.
Therefore, based on the analysis of packet loss probability and AoI, the time allocation ratio is limited from 0.1 to 0.4, while considering the size limit of radar information rate simultaneously.

\begin{figure}[!t]
    \centering
    \label{figa}
    \subfloat[]{\includegraphics[width=0.5\textwidth]{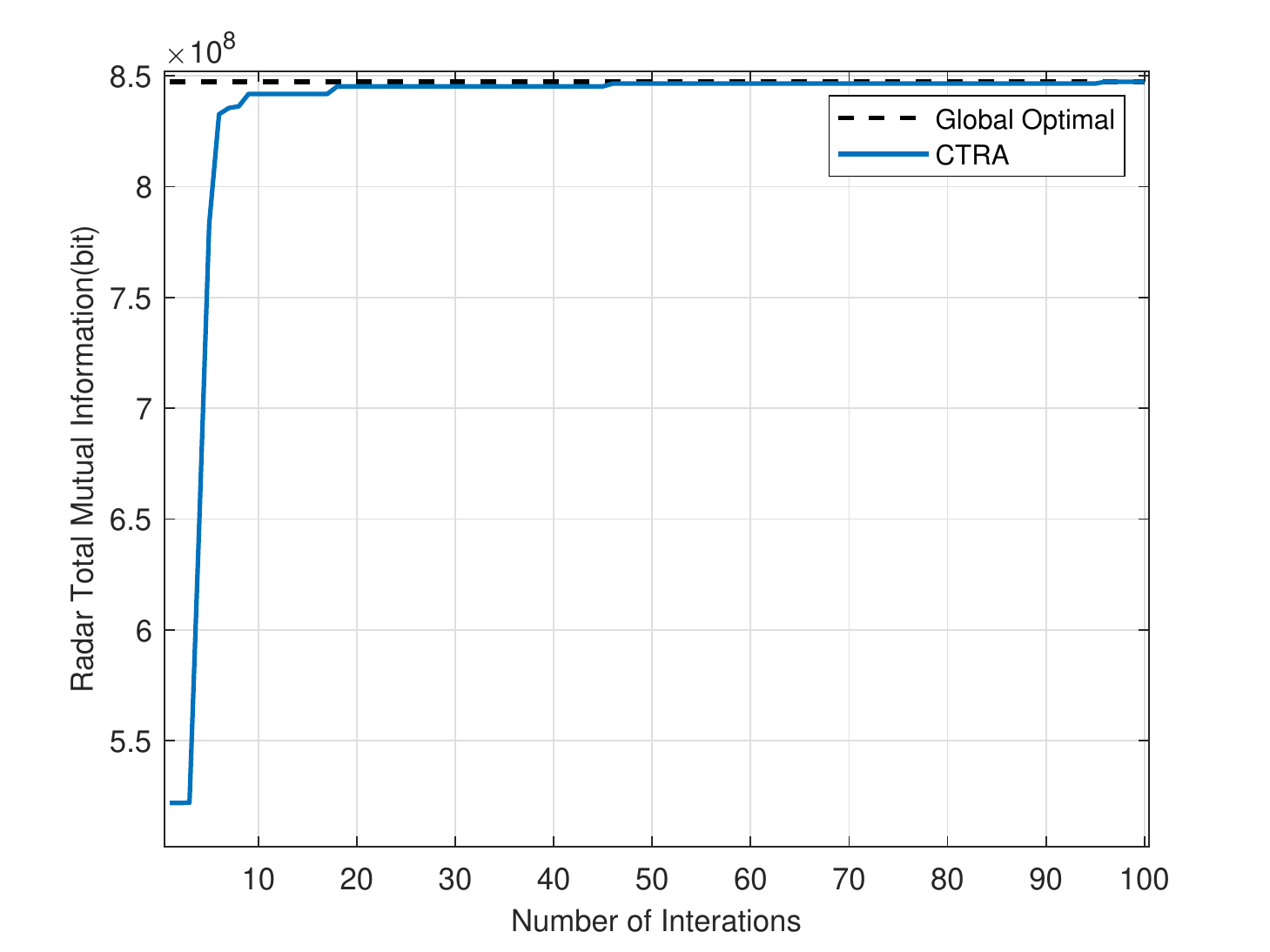}
    \label{figb}}
    \hspace{.1in}
    \subfloat[]{\includegraphics[width=0.5\textwidth]{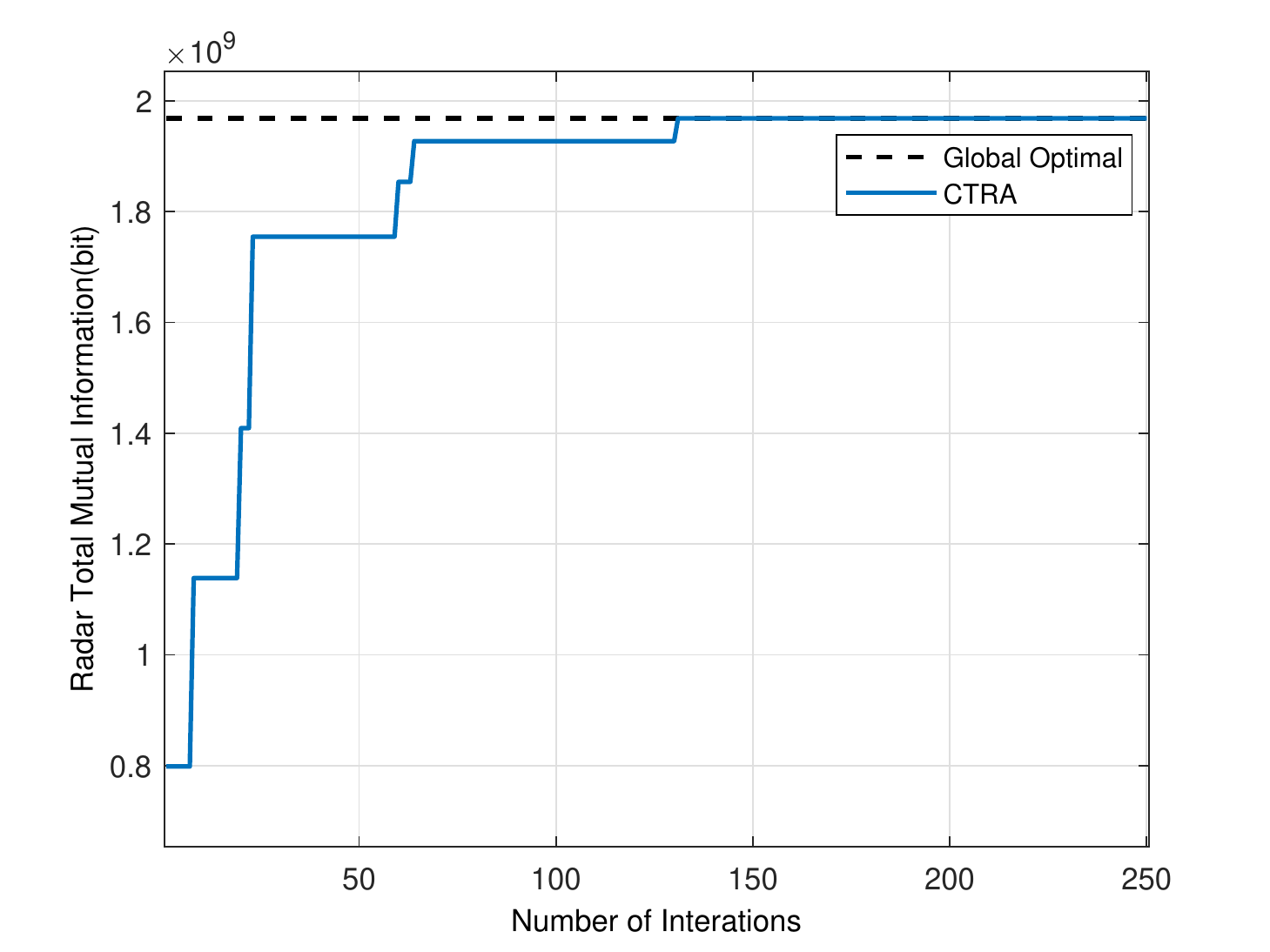}
    \label{fig2x}}
    \caption{Fig. 10. Convergence of proposed CTRA algorithm with different number of vehicles. (a) $N=5$, (b) $N=10$.}
    \label{CTRAconv}
\end{figure}
Fig. 10 shows the convergence performance of the proposed CTRA algorithm in the case of a small number of vehicles $N=5$, and a large number of vehicles $N=10$. The initial time allocation strategy of all vehicles is set to the minimum value ${a_m} = 0.1$ to ensure that a feasible optimal solution can be achieved. The simulation results show that the radar total MI increases with the increase of iterations. After a limited number of iterations, the system is stable, which can achieve a stable and feasible strategy. When the vehicle group size is large in Fig. 10(b), the convergence speed of the proposed CTRA algorithm is lower than that of the small vehicle group in Fig. 10(a), and the radar total MI is higher than that of the small vehicle group.

\begin{figure}[t]
\centerline{\includegraphics[width=0.5\textwidth]{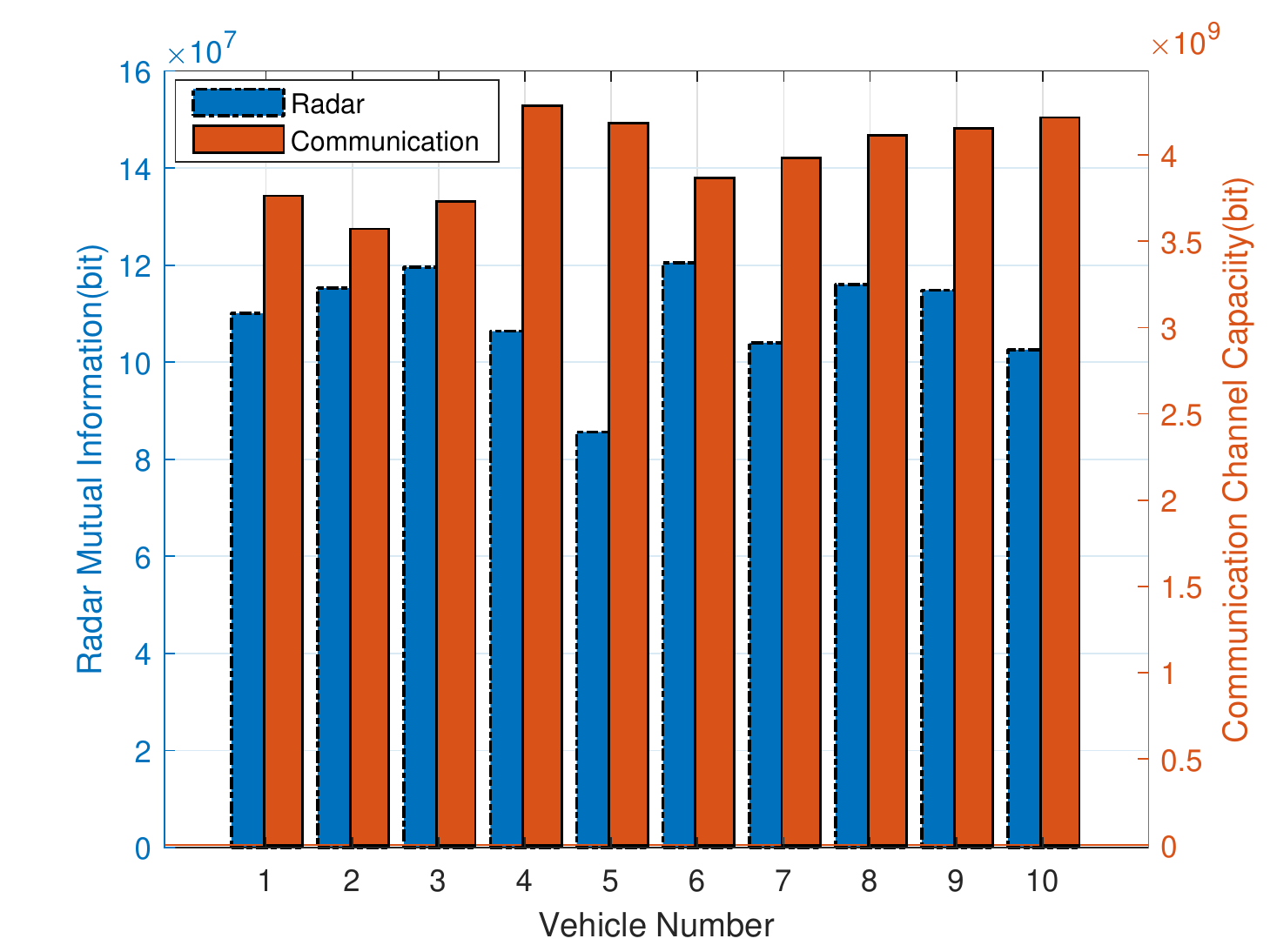}}
\caption{Fig. 11. The radar mutual information and communication channel capacity allocation using CTRA algorithm.}
\label{RATEper}
\end{figure}

Fig. 11 shows the configuration of the radar MI and the communication channel capacity of an isolated vehicle under the optimal allocation of time resources of the CTRA algorithm when $N = 10$. Although the radar MI and the communication channel capacity of each vehicle are different, the optimal time allocation scheme can meet the restrictions that the amount of radar MI of each vehicle is less than the amount of communication channel capacity, which means that the information detected by each vehicle's radar can be efficiently transmitted within the communication duration. And under the current mmWave system parameters, the communication channel capacity of an isolated vehicle can reach more than 3.5Gbits, and additional data can be transmitted according to the task requirements, which can support the high data rate transmission.

\begin{figure}[t]
\centerline{\includegraphics[width=0.5\textwidth]{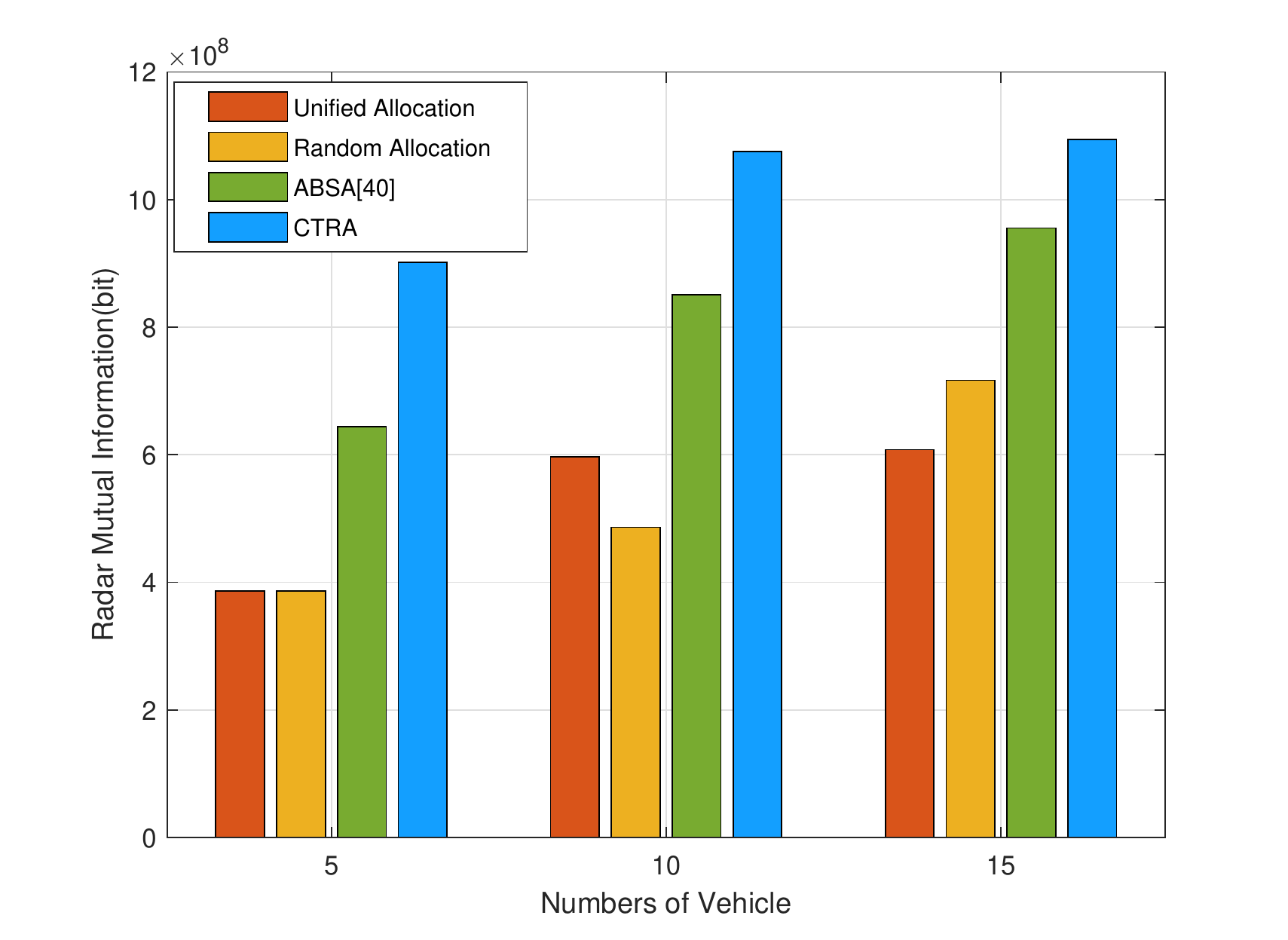}}
\caption{Fig. 12. Comparison of the radar total mutual information under different number of vehicles.}
\label{differNum}
\end{figure}

In order to verify the effectiveness and superiority of the CTRA algorithm for the time resource allocation of JSCIS, the performance comparison and evaluation of typical time resource allocation methods are performed, including the uniform time allocation method, the random time resource allocation strategy, the simulated annealing method (ABSA) [40], and the proposed CTRA algorithm in this paper. In the uniform time allocation method, each vehicle adopts the same duration allocation ratio, which is lack of flexibility. It is worth noting that the uniform allocation and random allocation strategies may result in a negative value for the final utility function. Furthermore, a time allocation strategy to make the radar total MI positive is selected for performance comparison in the simulation. Fig. 12 compares the performance of several typical time resource allocation methods for different numbers of vehicles, and evaluates the performance of different algorithms based on the radar total MI. The higher the radar total MI, the better the final utility. The results show that the CTRA algorithm has a better performance than other algorithms, which can achieve a higher radar total MI and improve the performance of communication system. The random allocation method has a large contingency, with no obvious advantages compared with the uniform allocation method. However, other algorithms using non-uniform time allocation are better than the uniform time allocation. The ABSA method can also improve system performance to some extent, but it is slightly inferior to the proposed CTRA algorithm. Specifically, compared with the ABSA method, the proposed CTRA algorithm can increase the radar total MI by 40$\%$, 26.29$\%$, and 14.53$\%$ when the number of vehicles is 5, 10, and 15, respectively.

\begin{figure}[t]
\centerline{\includegraphics[width=0.5\textwidth]{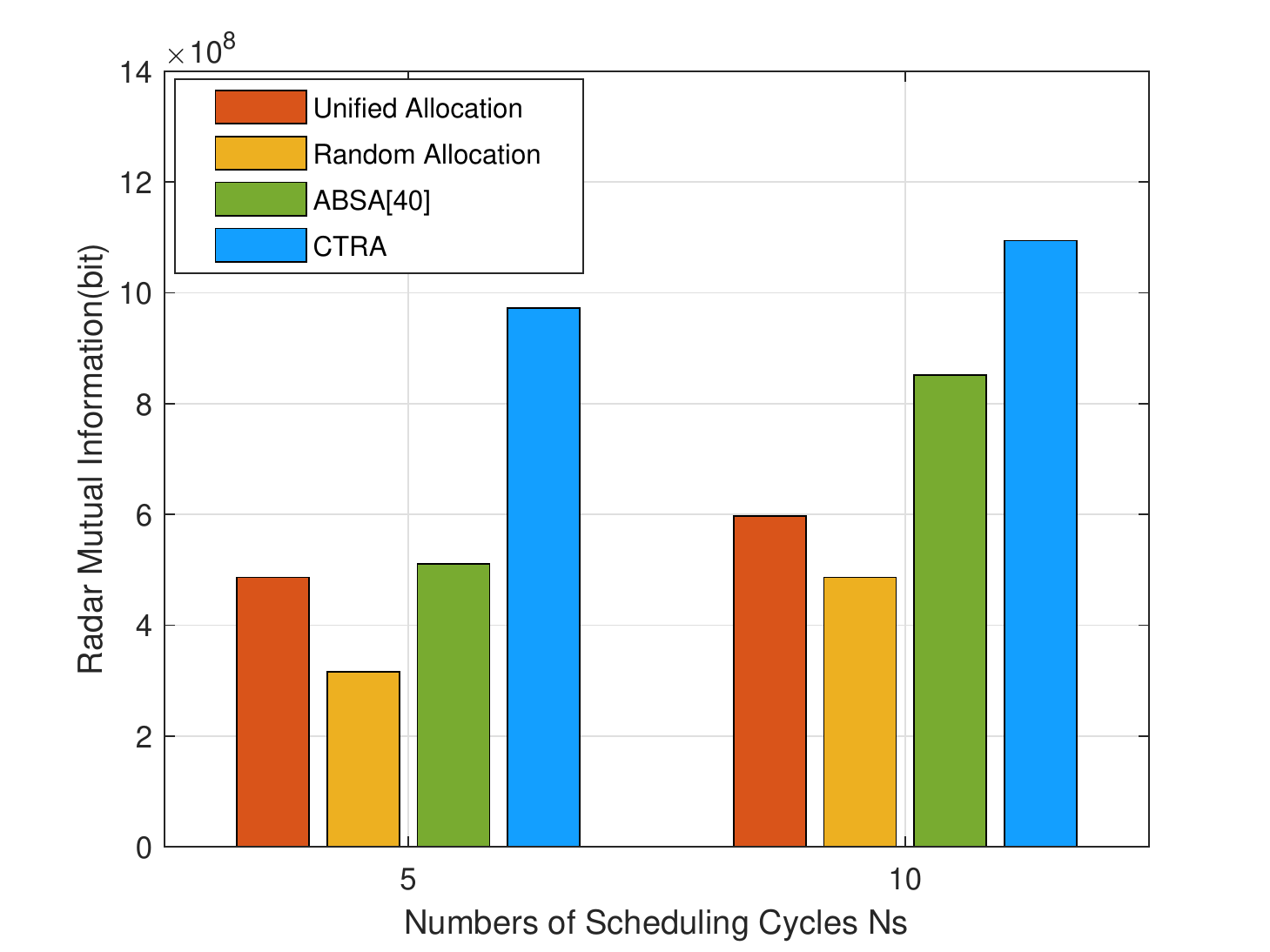}}
\caption{Fig. 13. Comparison of radar total mutual information under different scheduling cycles.}
\label{differNs}
\end{figure}

In order to compare the effect of different ${N_s}$ on various algorithms in a scheduling cycle, taking the practical frame structure of 10 ms into account, the number of subframes is set to 5 slots and 10 slots, respectively. As shown in Fig. 13, regardless of the value of ${N_s}$, excluding the random allocation, the radar total MI obtained by the non-uniform time allocation method is much better than the uniform allocation strategy. With the increase of ${N_s}$, the radar total MI obtained by different time allocation strategies increases. Therefore, the current system using ${N_s} = 10$ can obtain a better performance, making the overall system the highest utility.

\begin{figure}[t]
\centerline{\includegraphics[width=0.5\textwidth]{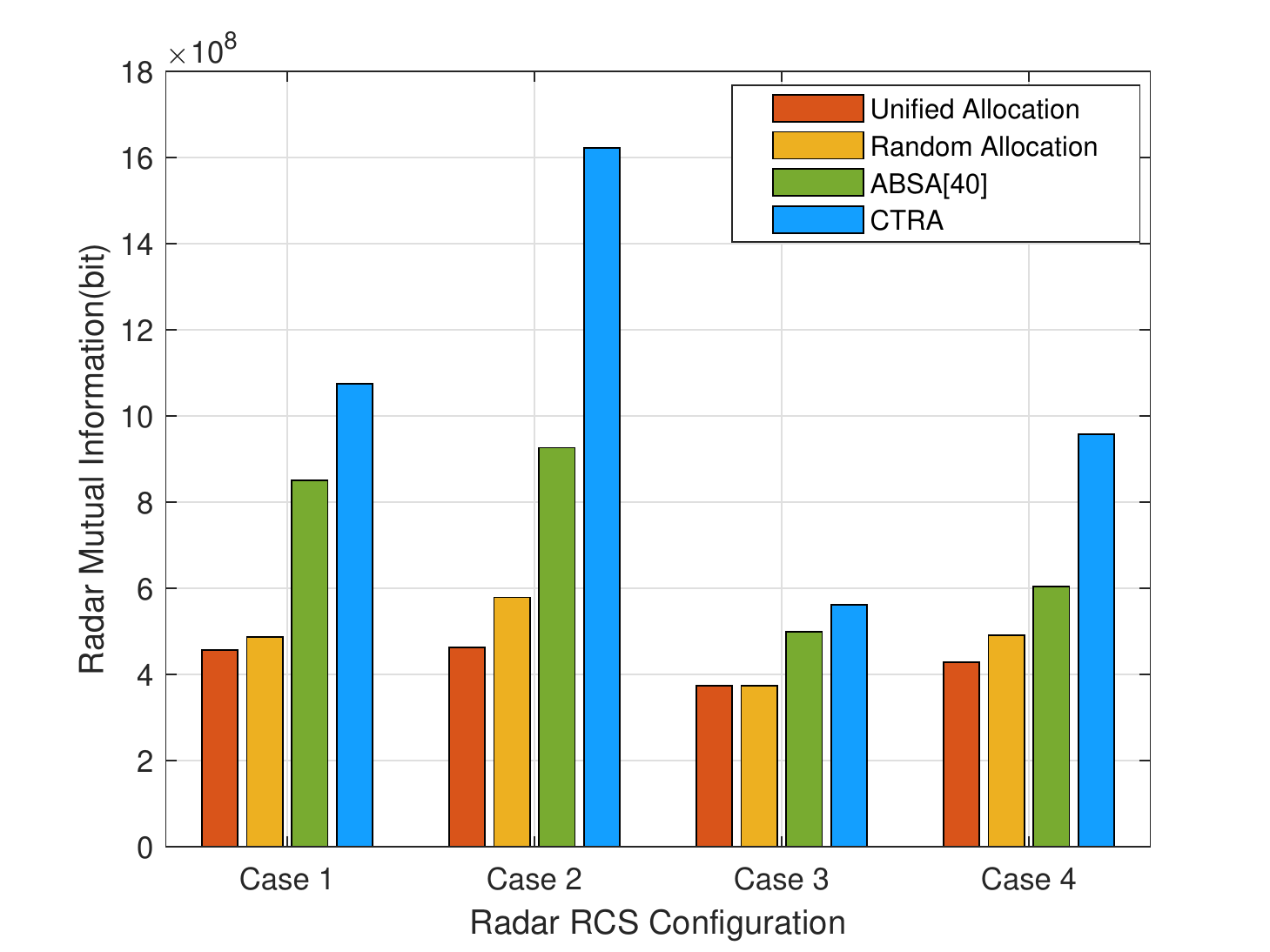}}
\caption{Fig. 14. Comparison of radar total mutual information under different radar RCS configurations.}
\label{differRCS}
\end{figure}

To evaluate the effect of different radar RCS configurations on the performance of various algorithms, four cases are considered as follows. Case 1 sets all ${\sigma ^{RCS}} = 1$; Case 2 sets the RCS of radar $i$ itself to ${\sigma ^{RCS}} = 1$, and the RCS between different radars is ${\sigma ^{RCS}} = 0.5$; Case 3 sets the RCS of radar $i$ itself to ${\sigma ^{RCS}} = 1$, and the RCS between different radars is ${\sigma ^{RCS}} = 2$; Case 4 sets the RCS of radar I itself to =1, and the RCS between different radars randomly takes a value from $0.5$ to $2$. The simulation results are shown in Fig. 14. All the algorithms in Case 2 can achieve the highest radar total MI, and the performance of the proposed CTRA algorithm is much better than other algorithms, while Case 3 has the lowest radar total MI. The performance of Case 1 and Case 4 are among the middle of all four cases, where Case 1 is slightly better than Case 4. Result shows that when the RCS of the current vehicle is smaller than that of other vehicles to the current vehicle $i$, the system impact on the current radar duration is also relatively small, and a larger radar total MI can be obtained. When the RCS of other vehicles on the current vehicle is large, radar interference will increase, and the radar total MI will be severely reduced. To sum up, in order to achieve a better performance, the impact of other radars on the current radar should be minimized in the practical implementation process.

\begin{figure}[!t]
    \centering
    \label{figa}
    \subfloat[]{\includegraphics[width=0.5\textwidth]{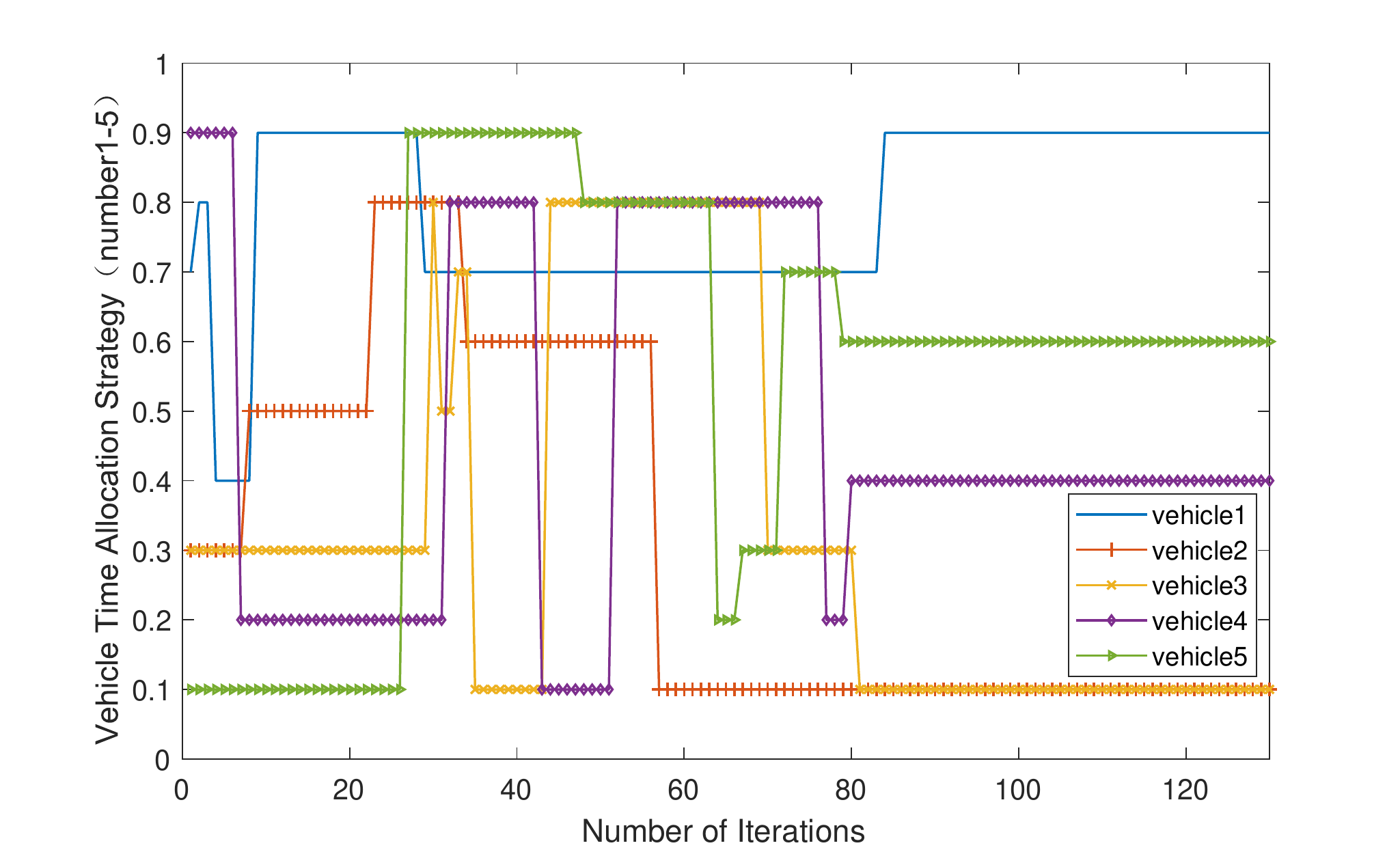}
    \label{figb}}
    \hspace{.1in}
    \subfloat[]{\includegraphics[width=0.5\textwidth]{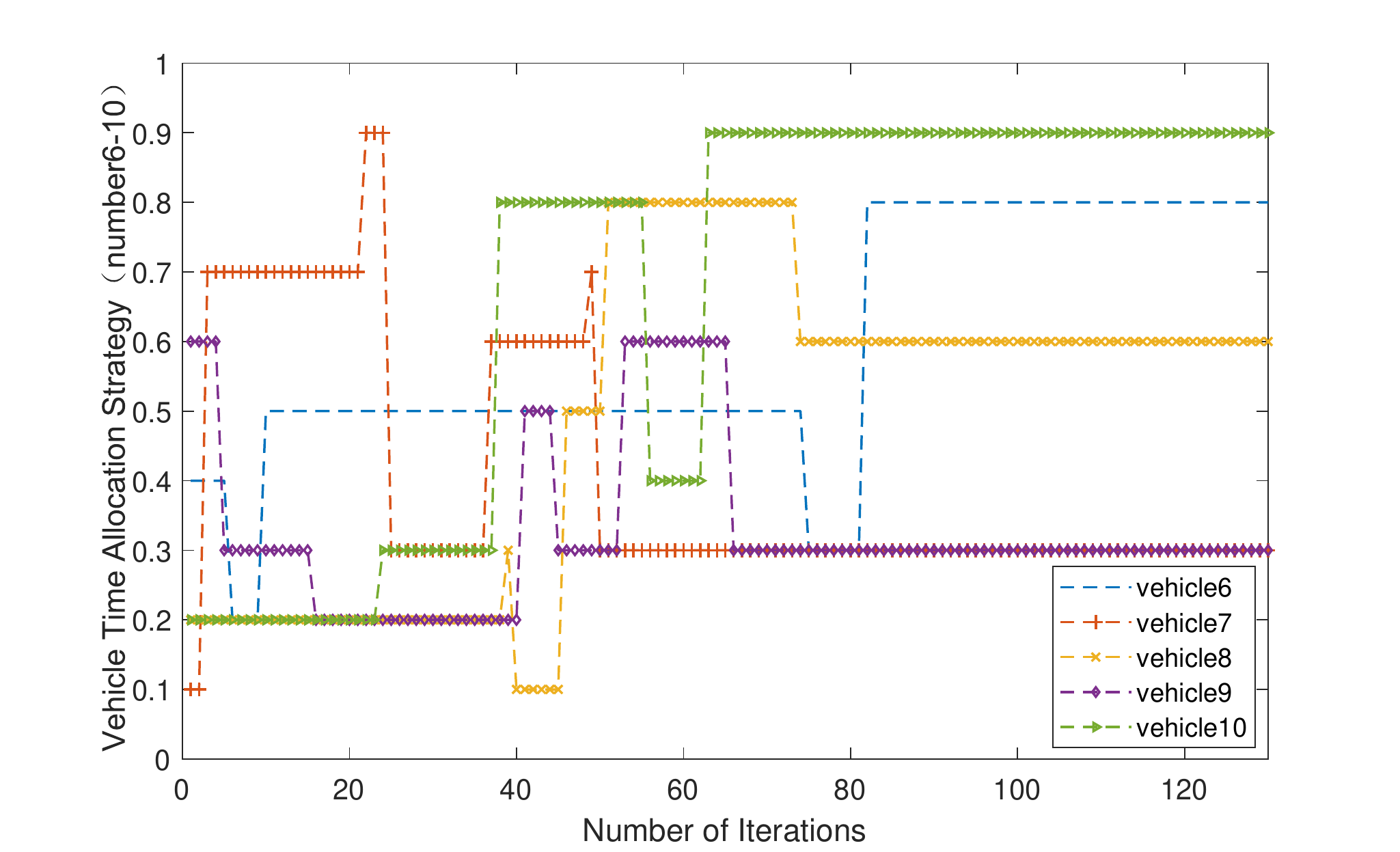}
    \label{fig2x}}
    \caption{Fig. 15. The status of changes of isolated vehicle's time resource allocation strategy with iterations. (a) Vehicle number 1-5, (b) Vehicle number 6-10.}
    \label{status}
\end{figure}

To analyze the convergence of the proposed CTRA, Fig. 15 shows the time resource allocation strategy of each vehicle with $N = 10$ as a function of the number of iterations. Due to the large number of vehicles, 10 vehicles are divided into two sub-graphs. It can be observed that the optimal time resource allocation strategy can be expressed as {0.9, 0.1, 0.1, 0.4, 0.6, 0.8, 0.3, 0.6, 0.3, 0.9} in the order of vehicle series 1-10. When the initial time is allocated as a minimum configuration, each vehicle is selected in turn for optimization, and the strategy is updated according to the proposed CTRA algorithm. As the number of iterations increases, each vehicle continuously optimizes its own time configuration. Finally, the CTRA algorithm is converged to the optimal time resource allocation strategy within 85 iterations.

%软件仿真结果分析的小结
In summary, the software simulation results prove that the best time duration allocation ratio for one vehicle can be achieved based on the analysis of queuing theory and AoI in Section II. In the multiple CAVs condition, the convergence of the proposed CTRA algorithm and the existence of the stable allocation ratio set for CAVs are proved. The proposed CTRA algorithm can ensure that all the radar sensing information can be efficiently transmitted in the communication duration. The performances of CTRA algorithm under different impact factors are also analyzed.

%B.硬件仿真结果分析
\subsection{Hardware Setup and Testbed Results Analysis}
%硬件测试综述
In order to realize the radar detection function in the perception phase of sensing and communication integrated system, the radar data frame is added to 5G NR mmWave communication frame. And the JSCIS hardware testbed is setup in the 28 GHz mmWave frequency band, which consists of two NI 5G mmWave communication platforms, two 64-elements phased array antennas and two horn antennas as shown in Fig. 16. The field test video of the proposed testbed is available in [41].

%硬件结构和实际测试平台场景
\begin{figure}[!t]
    \centering
    \label{figa}
    \subfloat[]{\includegraphics[width=0.5\textwidth]{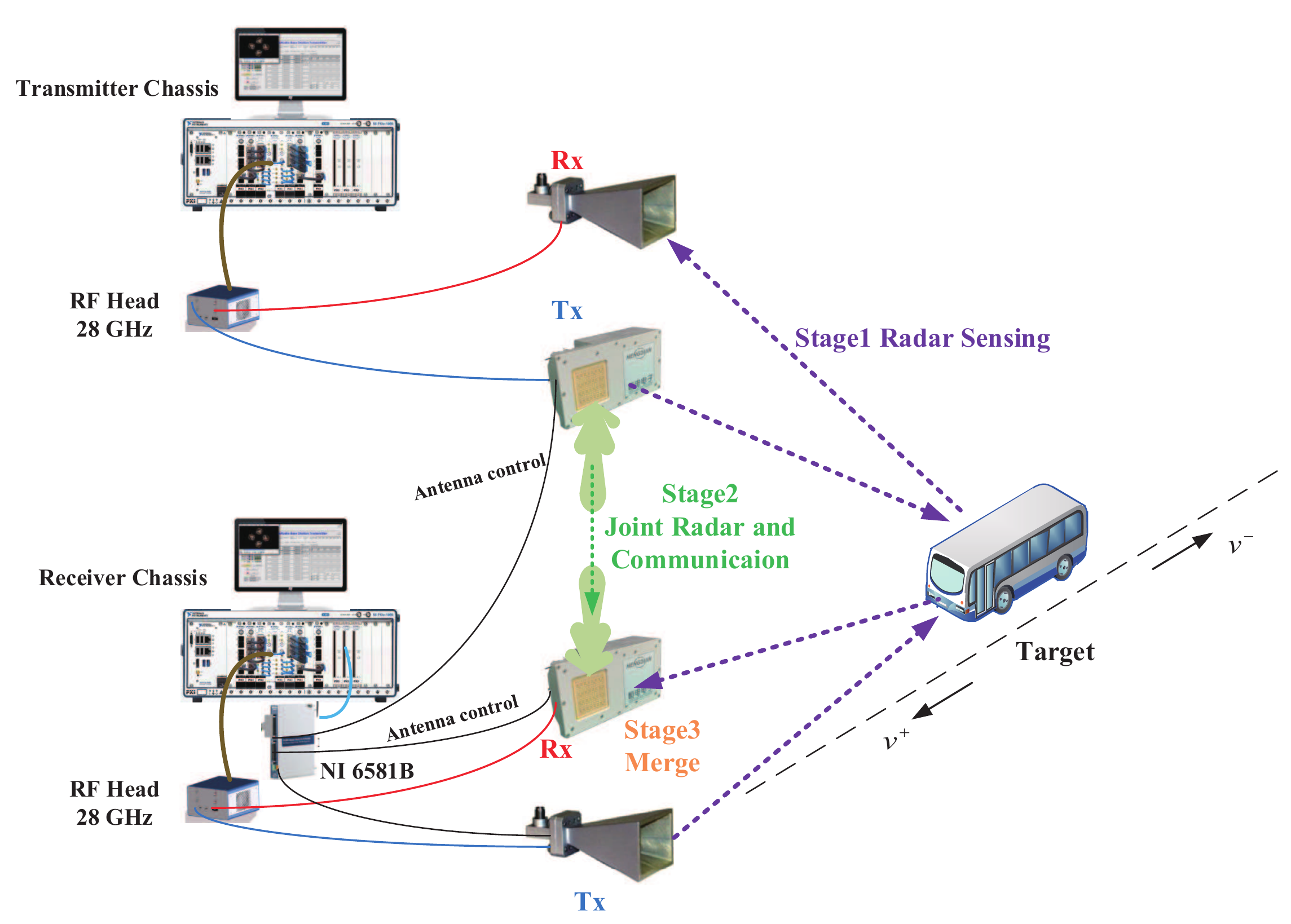}
    \label{figb}}
    \hspace{.1in}
    \subfloat[]{\includegraphics[width=0.5\textwidth]{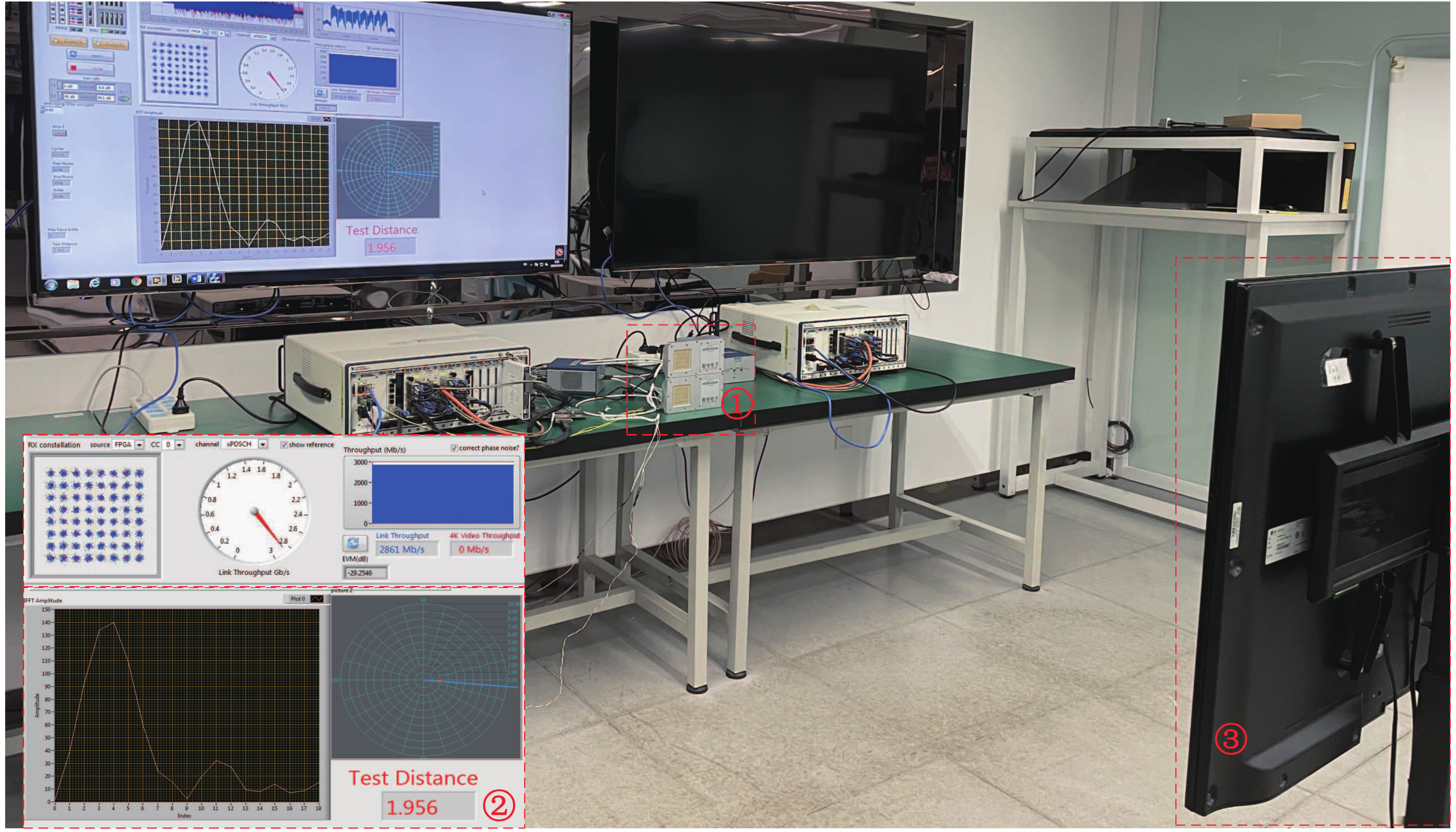}
    \label{fig2x}}
    \caption{Fig. 16. Hardware testbed of JSCIS system. (a) Hardware architecture of JSCIS system, (b) Field test scenario of JSCIS system. \ding{172} Phased array antenna HDTX270280-64CH. \ding{173} Test distance. \ding{174} Moving target.}
    \label{status}
\end{figure}

The hardware testbed architecture of JSCIS is shown in Fig. 16(a), where both horn antennas and phased array antennas are used at the transmitter and receiver sides. In stage 1, two JSCIS integrated equipments (IEs) are operating in the radar mode to detect the target independently. In stage 2, the IEs switch to the communication mode, establish one mmWave communication link from IE 1 to IE 2 via two phased array antennas, and transmit the raw sensing data of IE 1 in stage 1 to IE 2.

The hardware testbed is shown in Fig. 16(b). In the field test, one of the 5G mmWave platforms is set as a transceiver, and two phased array antennas are connected to Tx and Rx ends of the RF head as transmitter and receiver. The distance range is set from 0.5 m to 9 m. The radar detection data is compared with the actual range data to obtain the accuracy of the radar ranging in the JSCIS. Two phased array antennas \ding{172} are used as a fixed IE to send integrated detection signals. In the test, \ding{174} is used as the target with a moving speed of 1 m/s. The receiver performs 4096 points Inverse Fast Fourier Transform (IFFT) radar signal processing on the reflected echo, which can achieve the ranging radar performance with a resolution of 0.488 m. The refresh rate of radar data is 10 ms, and the system updates the detection data in the form of average value every 3 seconds, as shown in \ding{173}. When the target is placed at a distance of 2 m, the test result is 1.956 m, and the distance accuracy of target detection is $\pm$ 0.044 m.

%帧结构
\begin{figure}[t]
\centerline{\includegraphics[width=0.5\textwidth]{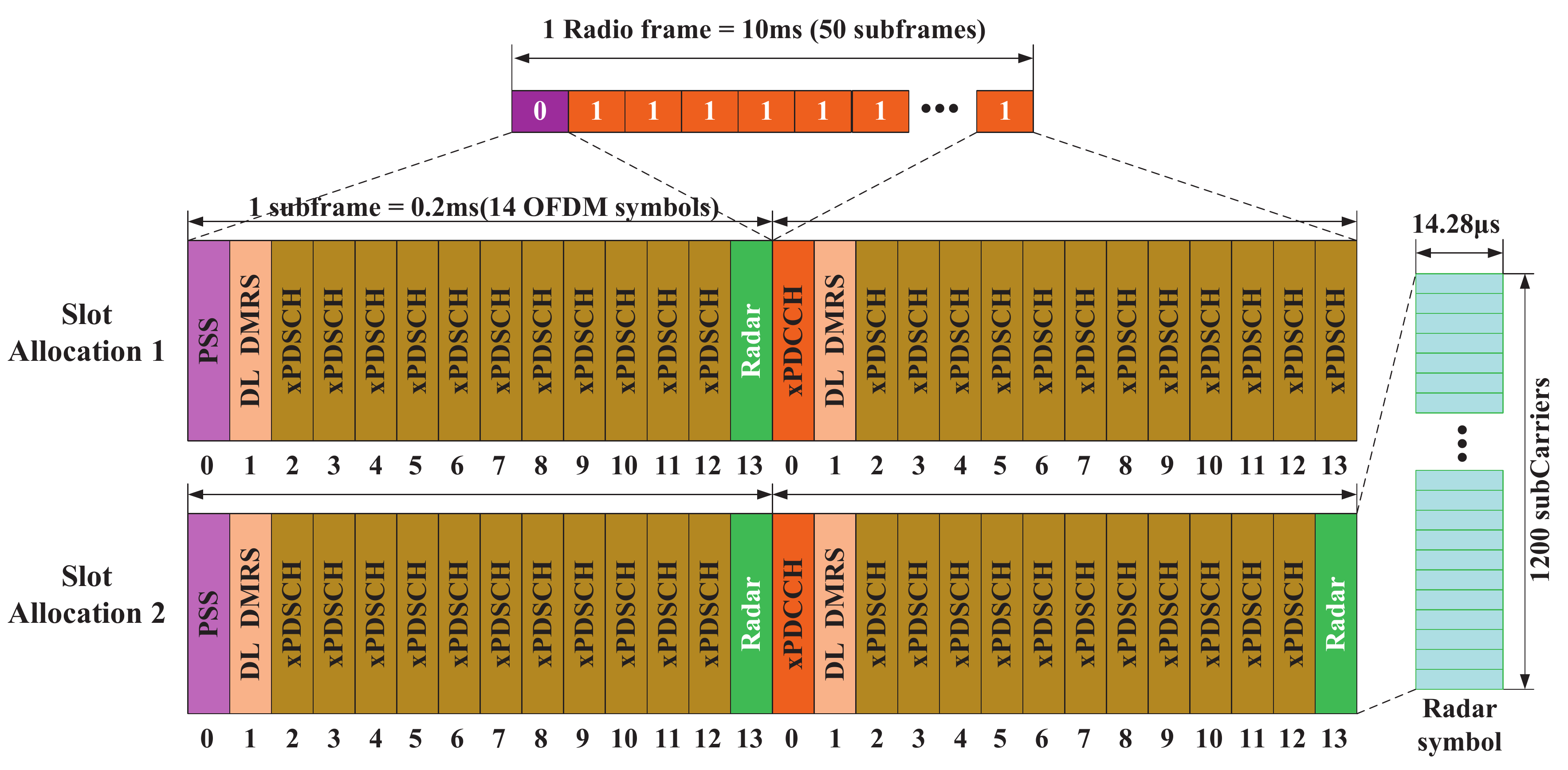}}
\caption{Fig. 17. Frame structure design of JSCIS system.}
\label{frame}
\end{figure}

Fig. 17 shows the slot allocation (SA) design of JSCIS. Each frame of JSCIS is 10 ms, which contains 50 subframes (SFs) in [42]. Each SF contains 14 Orthogonal Frequency Division Multiplexing (OFDM) symbols, and each symbol length is ${\rm{14}}{\rm{.28}} \mu s$. The SF type contains SF type0 and SF type1. Both Slot Allocation 1 (SA 1) and Slot Allocation 2 (SA 2) are shown in Fig. 17. Different slot allocation methods are realized by $R = K{}_R/\left( {K{}_R + {{\rm K}_C}} \right)$, where $K{}_R$ is the number of radar symbols and $K{}_C$ is the number of physical downlink shared channel (PDSCH) symbols in one radio frame. ${R_{\rm{1}}} = {\rm{1}}/\left( {{\rm{12}} \times {\rm{50}}} \right) = {\rm{1}}/{\rm{600}}$ in SA 1 denotes only one radar symbol is filled in SF 0. ${R_{\rm{2}}} = {\rm{50}}/\left( {{\rm{12}} \times {\rm{50}}} \right) = {\rm{50}}/{\rm{600}}$ in SA 2 denotes one radar symbol is added in SF 0 and SF 1. The number of subCarriers is 1200, and the filled 1200 valid data is used to complete the function of radar detection range in the process of JSCIS.

%雷达测距仿真结果
\begin{figure}[t]
\centerline{\includegraphics[width=0.5\textwidth]{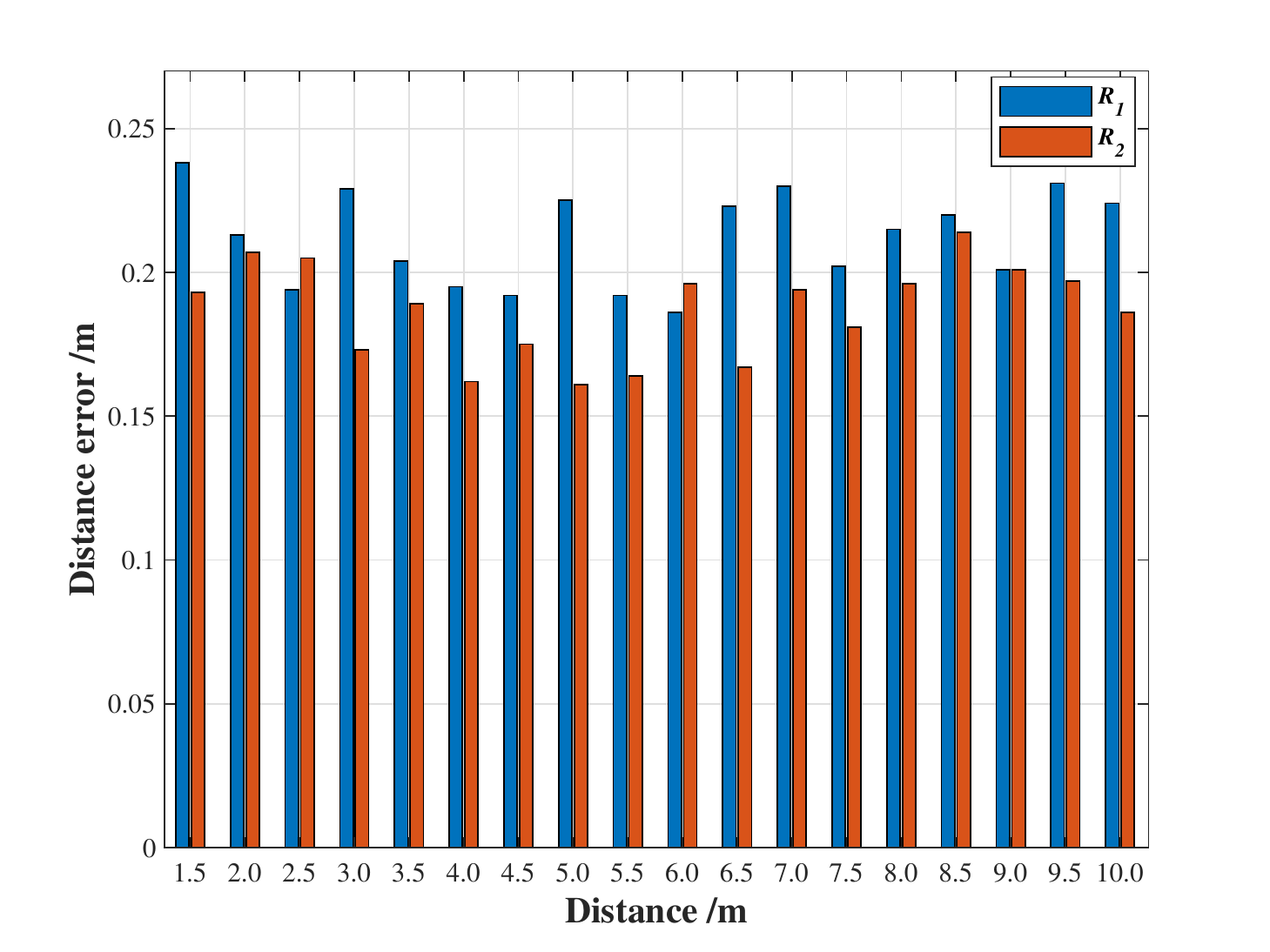}}
\caption{Fig. 18. Ranging results of JSCIS system.}
\label{ranging}
\end{figure}

Fig. 18 shows the radar test results of JSCIS under different time slot allocations with the target moving. The ranging accuracy is expressed in the form of distance error, where ${R_{\rm{1}}}$ and ${R_{\rm{2}}}$ correspond to SA 1 and SA 2 in Fig. 17, respectively. The target distance is set between 1.5 m and 10 m. The results show that the radar ranging accuracy of the integrated system is within $\pm$ 0.25 m, where the communication throughput of JSCIS can achieve a stable data rate of 2.8 Gbps. With the increase of radar time slot allocation ratio from ${R_{\rm{1}}}$ to ${R_{\rm{2}}}$, the radar ranging performance is also improved depicted by the decrease of distance error.

%第五部分：结论
\section{CONCLUSION}

In this paper, we have proposed the JSCIS to support the dynamic frame structure configuration for sensing and communication dual functions based on the 5G NR protocol in 28 GHz mmWave frequency band. And the best time duration allocation ratio of sensing and communication dual functions for one vehicle is achieved by modeling the $M/M/1$ queuing problem using the age of information (AoI) in this paper. Considering multiple CAVs, the resource allocation optimization problem has been formulated as a non-cooperative game, and the feasibility and existence of pure strategy NE are proved theoretically. Further, the CTRA algorithm is proposed to achieve the best feasible pure strategy NE. Both simulation and hardware testbed are designed and developed in this paper. The simulation results show that the proposed CTRA algorithm is convergent and can obtain the optimal system time allocation strategy more effectively than other algorithms. And the proposed CTRA algorithm can improve the radar total mutual information by 26$\%$. Finally, the hardware testbed results verify that the feasibility of the proposed JSCIS is achieved with an acceptable radar ranging accuracy within $\pm$ 0.25 m, as well as a stable data rate of 2.8 Gbps, which paves the way for the implementation of joint sensing and communication design in CAVs.

\bibliographystyle{IEEEtran}
\bibliography{IEEE,ISAC}

\ifCLASSOPTIONcaptionsoff
  \newpage
\fi

\end{document}